\documentclass[aps,prd,preprintnumbers,showpacs,showkeys,nofootinbib,
superscriptaddress,fleqn,floatfix,tightenlines,10pt,
byrevtex]{revtex4-2}
\usepackage{amsmath,amsfonts,amssymb,amscd,amsxtra,amsthm}
\usepackage{slashed}
\usepackage{appendix}
\usepackage[unicode]{hyperref}
\usepackage{graphicx,subfigure}
\usepackage{indentfirst}
\usepackage{float}
\usepackage{bm}
\usepackage{physics}
\usepackage{multirow}
\usepackage{makecell}
\usepackage{ltablex}
\usepackage{siunitx}

\usepackage[utf8]{inputenc}
\usepackage[normalem]{ulem} 
\usepackage[dvipsnames]{xcolor} 
\graphicspath{{./}}
\usepackage{indentfirst}
\usepackage{float}
\usepackage{cancel}

\usepackage{umoline}

\makeatletter

\begin{document}
\preprint{INHA-NTG-03/2026}
\title{Electromagnetic form factors of the nucleon from the
  instanton vacuum}    

\author{Hui-Jae Lee}
\email{hjlee6674@inha.edu}
\affiliation{Department of Physics, Inha University, Incheon 22212,
  Republic of Korea}
  
\author{Yongwoo Choi}
\email{sunctchoi@gmail.com}
\affiliation{Institute of Quantum Science, Inha University, Incheon 22212,
  Republic of Korea}  
\affiliation{The Center for High Energy Physics, Kyungpook National
  University, Daegu 41566, Republic of Korea} 

\author{Hyun-Chul Kim}
\email{hchkim@inha.ac.kr}
\affiliation{Department of Physics, Inha University, Incheon 22212,
  Republic of Korea} 
\affiliation{Institute of Quantum Science, Inha University, Incheon 22212,
  Republic of Korea} 
\affiliation{School of Physics, Korea Institute for Advanced Study
  (KIAS), Seoul 02455, Republic of Korea} 

\begin{abstract}
We investigate the electromagnetic form factors of the nucleon within
an effective chiral theory derived from the QCD instanton vacuum,
taking into account the finite current quark mass. The
momentum-dependent dynamical quark mass, generated by the
instanton-antiinstanton medium, naturally plays the role of a
regulator, so that no additional regularization is required to tame
the divergences arising from quark loops. The instanton parameters,
the average instanton size $\bar{\rho}=0.35$ fm and the average
interdistance $\bar{R}=0.86$ fm, together with the dynamical quark
mass at zero virtuality $M_0=385$ MeV, are all fixed by the
saddle-point equation beyond the chiral limit, leaving no adjustable
free parameter in the present calculation. We compute the Sachs
electric and magnetic form factors of the proton and neutron, the
nucleon charge and magnetization radii, the magnetic moments, and the
ratios $\mu_{p,n} G_E^{p,n}(Q^2)/G_M^{p,n}(Q^2)$. The present results
are compared with the experimental data, the chiral quark-soliton
model ($\chi$QSM), and the Kelly parametrization. The proton charge
radius, $\sqrt{\langle r^2 \rangle_\mathrm{ch}^p}=0.841$ fm, is in
remarkable agreement with the recent muonic-hydrogen value, and the
$Q^2$ dependence of the proton form-factor ratio
$\mu_p G_E^p/G_M^p$ is reproduced very well, in clear contrast to the
$\chi$QSM. The overall agreement with the experimental data confirms
that the effective chiral theory derived from the QCD instanton
vacuum provides a consistent and predictive framework for describing
the electromagnetic structure of the nucleon.
\end{abstract}

\date{\today}
\maketitle


\section{Introduction}
\label{sec:1}
Electromagnetic (EM) form factors are indispensable observables for
unraveling the internal structure of the
nucleon~\cite{Perdrisat:2006hj, Arrington:2011kb, Punjabi:2015bba}.
Since they encode the spatial distributions of the electric charge and
magnetization, these form factors represent fundamental properties of
the nucleon. The investigation of nucleon EM form factors has a long
history, dating back to the pioneering electron-scattering experiments
conducted by Hofstadter in the 1950s~\cite{Hofstadter:1956qs,
Yennie:1957skg}. The data on the neutron electric and magnetic form
factors have been steadily refined over the
decades~\cite{Bartel:1973rf, E94110:2004lsx, A1:2013fsc,
Meyerhoff:1994ev, Eden:1994ji, Becker:1999tw, Herberg:1999ud,
Ostrick:1999xa, Passchier:1999cj, Rohe:1999sh, E93026:2001css,
Bermuth:2003qh, JeffersonLabE93-026:2003tty, Glazier:2004ny,
BLAST:2008bub, E93-038:2003ixb, JeffersonLaboratoryE93-038:2005ryd,
Crawford:2006rz, Ron:2007vr, JeffersonLabHallA:2011yyi, Zhan:2011ji}.
The compilation of these data has enabled quantitative
phenomenological fits~\cite{Kelly:2004hm, Alberico:2008sz,
Qattan:2012zf}. In parallel, results from lattice calculations have
also been refined~\cite{Capitani:2015sba, Jang:2019jkn,
Alexandrou:2018sjm, Alexandrou:2025vto}. Nevertheless, despite decades
of extensive theoretical and experimental efforts, a complete
understanding of these properties remains elusive, and several aspects
are still subject to vigorous debate. A prominent example is the
proton charge radius puzzle~\cite{Pohl:2010zza, Antognini:2013txn,
Horbatsch:2015qda, Arrington:2015ria, Higinbotham:2015rja,
Fleurbaey:2018fih, Bezginov:2019mdi, Xiong:2019umf, Grinin:2020txk,
Atac:2021wqj, Brandt:2021yor, Scheidegger:2024rrm, Mohr:2024kco,
Maisenbacher:2026nau}, in which discrepancies between recent precision
measurements based on muonic and atomic hydrogen and the historical
data from electron-proton scattering have revealed open questions in
our understanding of the nucleon structure, calling for
continued theoretical investigation (see also the following
reviews~\cite{Pohl:2013yb, Gao:2021sml, Antognini:2022xoo}). The
Particle Data Group (PDG) illustrates how the average value of the
proton charge radius has evolved over the past two decades: in 2010,
the PDG average was given as $\sqrt{\langle
r^2\rangle_{\mathrm{ch}}^p}= 0.8768 \pm
0.0069\,\mathrm{fm}$~\cite{ParticleDataGroup:2010dbb}, whereas it is
now reported as $\sqrt{\langle r^2\rangle_{\mathrm{ch}}^p}= 0.8409\pm
0.0004\,\mathrm{fm}$~\cite{ParticleDataGroup:2024cfk}. 

In the meantime, the EM structure of the nucleon has been examined
from a more general perspective. The form factors of the nucleon are
now regarded as the first moments of the generalized parton distributions
(GPDs)~\cite{Muller:1994ses, Ji:1996nm, Radyushkin:1996ru,
Goeke:2001tz, Diehl:2003ny}. The first moments of the vector GPDs are
identified as the Dirac and Pauli form factors, which has also led to
discussions of the transverse charge distributions of the nucleon, in
which the significance of relativistic effects on the nucleon structure
has been emphasized~\cite{Miller:2007uy, Lorce:2020onh, Kim:2021kum}. Moreover,
the spin-orbit correlations in the nucleon are also deeply related to
the EM form factors of the nucleon~\cite{Leader:2013jra,
Lorce:2014mxa, Kim:2024cbq}. Considering that the spin-orbit
correlations originate from the twist-3 effective spin-orbit
correlation operator involving gluons, the EM structure of the nucleon
may be connected to gluonic effects, even though gluons do not carry
electric charge. Thus, precise information on the EM properties of the
nucleon may shed light on the intricate internal quark-gluon structure
of the nucleon.

In the present work, we therefore investigate the electromagnetic
form factors of the nucleon and related observables on the basis of
an effective chiral theory derived from the QCD instanton vacuum.
The instanton dilute liquid model was developed as a theoretical
framework for explaining the nonperturbative structure of the QCD
vacuum and the structure of low-lying
hadrons~\cite{Shuryak:1981ff,Diakonov:1983hh,Diakonov:1985eg} (see
also the reviews~\cite{Schafer:1996wv, Diakonov:2002fq}). The
instanton-antiinstanton medium is characterized by the average size
of the instanton $\bar{\rho}$ and the average interdistance between
instantons $\bar{R}$, which are approximately given by
$\bar{\rho}\approx 1/3$ fm and $\bar{R}\approx 1$ fm when one uses
$\Lambda_{\mathrm{QCD}}^{\overline{\mathrm{MS}}} = 280$
MeV~\cite{Shuryak:1981ff,Diakonov:1983hh,Diakonov:2002fq} (see also
Ref.~\cite{Liu:2025ldh} for a detailed discussion of the instanton
size). These values of $\bar{\rho}$ and $\bar{R}$ yield the correct
values of the gluon condensate~\cite{Shifman:1978bx} and of the
topological susceptibility of the vacuum~\cite{Diakonov:1995qy}.
Since we consider a finite current quark mass, we need to
reconstruct the instanton-antiinstanton medium beyond the chiral
limit. In Refs.~\cite{Kim:2005jc,Goeke:2007nc}, the saddle-point
equation was derived in the presence of the $1/N_c$ meson-loop
corrections and beyond the chiral limit, and the values of
$\bar{\rho}$ and $\bar{R}$ were readjusted to $\bar{\rho}=0.35$ fm
and $\bar{R}=0.86$ fm, with $f_\pi=88$ MeV and $\langle
i\psi^\dagger\psi \rangle_0 = (255\,\mathrm{MeV})^3$ required in the
chiral limit. The present theoretical framework can be regarded as an
effective theory for describing nonperturbative quark-gluon dynamics,
characterized by the low-energy normalization point given by the
inverse of the average instanton size,
$\bar{\rho}^{-1}\approx 0.6\,\mathrm{GeV}$. We refer to
Refs.~\cite{Diakonov:1985eg,Kim:1995bq,Diakonov:1995qy,Polyakov:2020cnc}
for a detailed discussion.

The instanton vacuum naturally describes the spontaneous breakdown of
chiral symmetry (SB$\chi$S): as a quark propagates through the
instanton-antiinstanton medium, it hops from one random instanton to
an antiinstanton. Its helicity changes during this process,
since the instanton zero mode has right-handed helicity whereas the
antiinstanton zero mode has left-handed helicity. Repeating this
process infinitely many times, the quark acquires a dynamical quark
mass $M(k)$, which is momentum-dependent~\cite{Diakonov:1985eg,
  Diakonov:2002fq}. The instanton vacuum then generates an
effective nonlocal $2N_f$ quark-quark interaction, where $N_f$ denotes
the number of flavors. Bosonizing this quark-quark interaction,
we obtain an effective chiral action with a momentum-dependent
dynamical quark mass. The value of the dynamical quark mass at
zero virtuality is determined by the saddle-point equation derived
from the instanton vacuum~\cite{Diakonov:1985eg, Diakonov:2002fq}. Once
the value of $\Lambda_{\mathrm{QCD}}$ is given, all the dynamical
parameters are fixed. Thus, we have no free parameters to adjust.
Musakhanov improved the formalism of Diakonov and Petrov by
incorporating the current quark mass beyond the chiral
limit~\cite{Musakhanov:1998wp, Musakhanov:2002vu}.

The effective chiral theory of the nucleon that we
employ for computing the nucleon EM form factors in this work is based
on this effective nonlocal chiral action~\cite{Choi:2025xha}.
The local version of this effective chiral theory is known as the
chiral quark-soliton model ($\chi$QSM)~\cite{Diakonov:1987ty,
  Wakamatsu:1990ud, Blotz:1992pw} (see also the following
reviews~\cite{Christov:1995vm, Diakonov:1997sj}). This
effective chiral theory of the nucleon was motivated by Witten's
pion mean-field idea~\cite{Witten:1979kh, Witten:1983tw,
  Witten:1983tx}. In the pion mean-field picture, the nucleon can be
viewed as a bound state of $N_c$ (the number of colors) valence
quarks in the large $N_c$ limit. The presence of these valence quarks
polarizes the vacuum self-consistently, and the vacuum polarization
produces the pion mean field that binds the $N_c$ valence quarks.
This self-consistent interaction stabilizes the entire system
consisting of the valence quarks and the Dirac continuum, and for
this reason it has also been called a non-topological chiral
soliton~\cite{Christov:1995vm}.
We will use this effective chiral theory as our theoretical framework
to compute the EM form factors of the nucleon. Since the
momentum-dependent dynamical quark mass plays the role of a regulator,
we do not need to introduce any additional regularization to tame the
divergences arising from quark loops. In addition, we also take into
account the current quark mass~\cite{Musakhanov:1998wp, Musakhanov:2002vu},
since the isovector charge and magnetization radii of the nucleon
diverge in the chiral limit~\cite{Adkins:1983hy}. 

This work is organized as follows. In Sec.~\ref{sec:2}, we define the
electromagnetic current and express its nucleon matrix element, which
is parametrized in terms of the Dirac and Pauli form factors of the
nucleon. We also relate these two form factors to the Sachs-type EM
form factors. Section~\ref{sec:3} explains the effective chiral theory
of the nucleon based on the instanton vacuum. We recapitulate the
formalism developed in Ref.~\cite{Choi:2025xha}.
We then show how the nucleon matrix element of the EM
current can be computed within the present theoretical
framework, and we derive the final expressions for the EM form factors
of the nucleon. In Sec.~\ref{sec:4}, we present the numerical results
for the EM form factors of the nucleon, and we compare them with the
experimental data. We also compute the nucleon charge and
magnetization radii, and present the results for the ratios 
of the electric and magnetic form factors. 
The final section is devoted to the summary and conclusions.
\section{Electromagnetic form factors of the nucleon}
\label{sec:2}
The EM quark current in flavor SU(2) is defined as
\begin{align}
  J_\mu(x) = \bar{\psi}(x)\gamma_\mu \hat{Q} \psi(x),
\label{eq:1}
\end{align}
where $\psi(x)$ denotes the quark field and $\gamma_\mu$ are the Dirac
gamma matrices. $\hat{Q}$ stands for the quark charge matrix, defined as
\begin{align}\label{eq:2-1}
  \hat{Q} = \begin{pmatrix}
    \frac{2}{3} & 0 \\
    0 & -\frac{1}{3}
  \end{pmatrix}
   = \frac{1}{6}\bm{1} + \frac{1}{2}\tau^3,
\end{align}
where $\tau^3$ is the third component of the Pauli matrices.
Accordingly, the EM current can be decomposed into the isoscalar and
isovector vector currents:
\begin{align}
J_\mu = J_\mu^{(0)} + J_\mu^{(3)},
\label{eq:3}
\end{align}
with
\begin{align}
  J_\mu^{(0)} = \frac{1}{6} \bar{\psi}(x)\gamma_\mu \psi(x),\;\;\;
  J_\mu^{(3)} = \frac{1}{2} \bar{\psi}(x)\gamma_\mu \tau^3 \psi(x).
\end{align}
The nucleon matrix element of the EM current can be
parametrized in terms of two real EM form factors: 
\begin{align}
  \bra{N(p',S')}J_\mu(0)\ket{N(p,S)} &= \bar{u}_N(p',S')\Big[
    \gamma_\mu F_1(Q^2) + \frac{i\sigma_{\mu\nu}\Delta^\nu}{2M_N}
  F_2(Q^2)  \Big]u_N(p,S) \cr
  &= \bar{u}_N(p',S')\Big[   \frac{M_NP^\mu}{P^2} G_E(Q^2)
    + \frac{i\epsilon^{\mu\nu\alpha\beta} \Delta_\nu P_\alpha
    \gamma_\beta \gamma_5}{2P^2} G_M(Q^2)
    \Big]u_N(p,S),
\label{eq:5}
\end{align}
where $M_N$ is the nucleon mass, $\Delta^\mu=(p'-p)^\mu$ is the 
spacelike four-momentum transfer, and $P^\mu=(p+p')^\mu/2$ 
represents the average nucleon momentum. $Q^2$ is defined as the
positive-definite momentum transfer squared, $Q^2=-\Delta^2$ with
$Q^2>0$. $u_N(p,S)$ denotes the Dirac spinor for the nucleon state
with momentum $p$ and spin $S$. $F_1(Q^2)$ and $F_2(Q^2)$ are
called the Dirac and Pauli form factors, respectively,
whereas $G_E(Q^2)$ and $G_M(Q^2)$ are known as the electric and
magnetic Sachs form factors, respectively~\cite{Sachs:1962zzc}. They  
are related to the Dirac and Pauli form factors as follows:
\begin{align}
  G_E(Q^2) &= F_1(Q^2) - \frac{Q^2}{4M_N^2}F_2(Q^2), \\
  G_M(Q^2) &= F_1(Q^2) + F_2(Q^2).
\end{align}
The Sachs form factors $G_E(Q^2)$ and $G_M(Q^2)$
correspond to the temporal and spatial components
of the nucleon matrix element, respectively, as follows:
\begin{align}\label{eq:2-2}
  \bra{N(p',S')}J^0(0)\ket{N(p,S)}
  &= \delta_{S'_3S_3} G_E(Q^2), \\
  \bra{N(p',S')}J^i(0)\ket{N(p,S)}
  &= \frac{i}{2M_N}\varepsilon^{ijk}\langle\sigma^j
    \rangle_{S'_3S_3}\Delta^k G_M(Q^2),
\end{align}
where $\langle\sigma^k\rangle_{S'_3S_3}$ denotes
the expectation value of the Pauli matrices with respect to
the initial and final spin states of
the nucleon. According to the definition of the isoscalar and
isovector EM current given in Eq.~\eqref{eq:3}, the Sachs form factors
can be decomposed into the isoscalar and isovector ones,
\begin{align}
  \label{eq:10}
  G_{E(M)}^{N} (Q^2) = \frac{1}{2}\left(G^{T=0}_{E(M)}(Q^2)
  + \tau_3 G^{T=1}_{E(M)}(Q^2) \right),
\end{align}
where $\tau_3$ corresponds to the eigenvalue for the third component of
the isospin Pauli matrix. Thus, the proton and neutron electric form
factors in the forward limit correspond to their charges 
\begin{align}
  \label{eq:2-4}
  G^p_E(0) = 1,\,\,\,
  G^n_E(0) = 0,
\end{align}
where the superscripts $p$ and $n$ denote
the proton and neutron, respectively. The
squared charge radii are defined as
the slopes of the nucleon electric form factors at $Q^2=0$
\begin{align}\label{eq:2-5}
  \langle r^2\rangle_E^{p(n)} =
  -6\frac{d G^{p(n)}_{E}(Q^2)}{dQ^2}
  \Bigg|_{Q^2=0}.
\end{align}
Similarly, the proton and neutron magnetic form factors in the forward
limit yield the corresponding magnetic moments,
\begin{align}\label{eq:2-6}
  \mu_{p(n)} = G_M^{p(n)}(0).
\end{align}
Finally, the squared magnetization
radii are defined as follows:
\begin{align}\label{eq:2-7}
  \langle r^2 \rangle_M^{p(n)} =
  -\frac{6}{\mu_{p(n)}}
  \frac{d G^{p(n)}_{M}(Q^2)}{dQ^2}
  \Bigg|_{Q^2=0}.
\end{align}
In this work, we will compute the Sachs EM form factors of the nucleon
within the framework of an effective chiral theory derived from the instanton
vacuum.
\section{Effective chiral theory from the QCD instanton vacuum}
\label{sec:3}
We start from the effective low-energy QCD
partition function obtained after bosonization of the $2N_f$ quark
interactions~\cite{Diakonov:1985eg,
  Diakonov:2002fq,  Musakhanov:1998wp, Musakhanov:2002vu}:
\begin{align}
\mathcal{Z}_{\mathrm{eff}}[\psi^\dagger,\psi, \pi^a] =& \int
\mathcal{D}\pi^a \int  \mathcal{D}\psi^\dagger  \mathcal{D}\psi                  \,\exp\left[
\int d^4 x \left\{ \psi_f^\dagger (x) (i\rlap{/}{\partial}) \psi^f (x)
  +i \psi_f^\dagger (x) m_f\psi^f (x)  \right. \right. \cr
 & \left.\left. +\, i M_0f(m_f) \int
  \frac{d^4k d^4 l}{(2\pi)^8}e^{i(k-l)\cdot x}~ F(k)F(l)
  \psi_f^{\dagger}(k)
  \left(U_g^f(x) P_L + U_g^{\dagger f}(x) P_R\right)
  \psi^g(l)    \right\}\right],
  \label{eq:15}
\end{align}
where $\pi^a$ represent the pseudo-Nambu-Goldstone (pNG) boson
fields, $\psi_f(x)$ is the quark field with flavor $f$, and
$m_f$ denotes the current quark mass
with flavor $f$. We assume isospin symmetry
($m_{\mathrm{u}}=m_{\mathrm{d}}$), and we define the average
current quark mass of the up and down quarks as
$\overline{m}=(m_{\mathrm{u}} + m_{\mathrm{d}})/2$.
The quark form factor $F(k)$ arises from the Fourier
transform of the fermionic zero mode in the instanton background
field:
\begin{align}
  F(k)
  = -z\frac{d}{dz}\Big[I_0(z)K_0(z)-I_1(z)K_1(z)\Big]
  \bigg|_{z=\frac{\bar{\rho}k}{2}},
\end{align}
where $I_n(x)$ and $K_n(x)$ denote the modified Bessel functions of
the first and second kinds, respectively. Owing to this
form factor, the dynamical quark mass $M(k)$ becomes
momentum-dependent. The function $f(m_f)$ originates from the $m_f$
dependence of the dynamical quark mass~\cite{Musakhanov:2002vu}:
\begin{align}
  f(m_f) = \sqrt{1+\frac{m_f^2}{d^2}}-\frac{m_f}{d},
\end{align}
where $d\approx 277$ MeV. Note that $d$ is naturally determined once 
the values of $\bar{\rho}$ and $\bar{R}$ are known.
The chiral field $U(x)$ is defined as $U (x) := \exp[i\pi^a(x)
\tau^a]$ with $a=1,\,2,\,3$. $P_L$ and $P_R$ stand for the left-handed
and right-handed chirality projection operators,
expressed as $P_L = (1-\gamma_5)/2$ and $P_R = (1+\gamma_5)/2$,
respectively. The combination $U(x) P_L + U^\dagger (x) P_R$ can be written
in the following compact form:
\begin{align}
U^{\gamma_5} (x) := \exp(i\pi^a \tau^a
\gamma_5) = U(x) P_L + U^\dagger (x) P_R.
\end{align}
Integrating over the quark fields, we obtain the following effective
partition function:
\begin{align}
\mathcal{Z}_{\mathrm{eff}}[\pi^a] =  \int \mathcal{D}\pi^a ~\mathrm{Det}\left[
  i\rlap{/}{\partial} +  i\overline{m} \bm{1} + i M_0 f(\overline{m})
  \overleftarrow{F}(i\partial) U^{\gamma_5}[\pi^a(x)]
    \overrightarrow{F}(i\partial)   \right].
\label{eq:19}
\end{align}
In the large $N_c$ limit, the pionic quantum fluctuations are
suppressed by $1/N_c$. Thus, we can integrate over the $\pi^a$ fields
in Eq.~\eqref{eq:19} around the classical saddle
point~\cite{Witten:1979kh, Witten:1983tw, Witten:1983tx}, which
yields the nonlocal effective chiral action (NE$\chi$A)
\begin{align}
S_{\mathrm{eff}}[U]= -N_c \mathrm{Tr}\log (D[U]),
\label{eq:20}
\end{align}
where the Dirac differential operator $D[U]$ is given by
\begin{align}
  D [U] =  i\rlap{/}{\partial} + i \overline{m} + i M_0 f(\overline{m})
  \overleftarrow{F}(i\partial) U^{\gamma_5} (x)
  \overrightarrow{F}(i\partial).
\label{eq:21}
\end{align}
$H[U]$ is the single-particle Dirac Hamiltonian
written as
\begin{align}\label{eq:22}
  H[U] = \gamma_4\gamma_i\partial_i+ \gamma_4 \overline{m}
  + M_0 f(\overline{m})
  \overleftarrow{F}\gamma_4U^{\gamma_5}\overrightarrow{F}.
\end{align}
In the present mean-field approach, the nucleon consists of
$N_c$ valence quarks~\cite{Diakonov:1987ty, Christov:1995vm,
  Diakonov:1997sj}, which are described by the quark propagator in
the background pion field~\cite{Choi:2025xha}
\begin{align}
  G_{f_{i}g_{i}}(\bm{0}, +\mathcal{T}/2|\bm{0},-\mathcal{T}/2) &=
\langle \bm{0}, +\mathcal{T}/2, f_i |
  D[U]^{-1}i\gamma_{4}| \bm{0}, -\mathcal{T}/2, g_i\rangle
= \int \frac{d\omega}{2\pi}~
\left\langle \bm{0}, f_i \left| \frac1{-i\omega+H} \right|
  \bm{0},g_i\right \rangle  e^{-i\omega \mathcal{T}},
\end{align}
where the single-particle Dirac Hamiltonian can be diagonalized as
$H|n_\omega\rangle = E_n(\omega)|n_\omega\rangle$. 
Note that the eigenenergies are given as functions of $\omega$. 
The detailed formalism for constructing the classical nucleon within 
this effective chiral theory can be found in Ref.~\cite{Choi:2025xha}.

To investigate the EM form factors of the nucleon within the present
effective chiral theory, we have to introduce the EM current into the
low-energy effective QCD partition function in Eq.~\eqref{eq:15}. However,
the nonlocal interaction arising from the momentum-dependent quark
mass breaks the U(1) gauge symmetry~\cite{Chretien:1954we}. Many
efforts have been devoted to resolving this problem~\cite{Chretien:1954we,
  Holdom:1989jb, Holdom:1990iq, Bos:1991sz, Holdom:1992fn,
  Ball:1993ak, Bowler:1994ir, Golli:1998rf, Broniowski:1999bz}.
As pointed out by Pobylitsa~\cite{Pobylitsa:1989uq}, the breakdown of
the gauge symmetry in the QCD instanton vacuum is deeply rooted in the
treatment of higher-order contributions to hadronic correlation functions in the
$N/V N_c$ expansion, where $N/V$ denotes the density of the instanton
medium. However, it is technically formidable to deal with such higher-order
corrections. Musakhanov and Kim~\cite{Musakhanov:2002xa}
introduced the external EM field into the QCD action and kept the gauge
invariance intact by introducing the gauge connection in the course of
constructing the effective $2N_f$ quark interactions. In the weak photon
field, it was shown that the gauge invariance can be preserved by
replacing the ordinary derivative with the covariant
one~\cite{Musakhanov:2002xa}, which is identical to the well-known
minimal substitution. This method has been successfully applied to many
observables~\cite{Kim:2004hd, Nam:2006sx, Nam:2007fx, Nam:2007gf,
  Goeke:2007nc, Son:2015bwa, Shim:2017wcq, Shim:2018rwv}. In
Ref.~\cite{Choi:2025xha}, the conservation of the baryon number was
demonstrated using this method.
Following Ref.~\cite{Musakhanov:2002xa}, we can express the gauged
effective low-energy partition function as
\begin{align}
  Z_{\mathrm{eff}}[\pi^a,v] =  \int \mathcal{D}\pi^a
  ~\exp(-S_{\mathrm{eff}}[\pi^a, v]),
\label{eq:24}
\end{align}
where the gauged NE$\chi$A is given by
\begin{align}
S_{\mathrm{eff}}[U,v] =-N_c\mathrm{Tr}\log \left[D(U,v)\right]
\label{eq:25}
\end{align}
with
\begin{align}
D(U,v)=  i\slashed{\nabla} + i\overline{m} + i M_0 \overleftarrow{F}(i\nabla)
  U^{\gamma_5}(x)   \overrightarrow{F}(i\nabla).
\end{align}
Here, $\nabla$ denotes the covariant derivative
\begin{align}
\nabla_\mu = \partial_\mu - i \mathcal{V}_\mu,
\end{align}
where $\mathcal{V}_\mu$ is the external photon field defined as
$\mathcal{V}_\mu = \hat{Q} v_\mu$.

To compute a nucleon matrix element in the present work, we have to
define the nucleon state. In the effective chiral theory, the nucleon
state consists of $N_c$ valence quarks, which can be expressed as
\begin{align}
\ket{N(p,S)}&=\mathcal{N}^*(p)\lim_{x_{4}\to-\infty}e^{ ip _{4}x_{4}}
\int\;d^{3}x\;e^{ i\bm{p }\cdot\bm{x}}J_{N}^{\dagger}(x)\ket{0},\cr
\bra{N(p',S')}&=\mathcal{N}(p')\lim_{y_{4}\to+\infty}e^{-ip'_{4}y_{4}}
\int\;d^{3}y\;e^{-i\bm{p'}\cdot\bm{y}}\bra{0}J_{N}(y),
\end{align}
where $\mathcal{N}^*$ and $\mathcal{N}$ are the normalization
constants of the nucleon state.
The Ioffe-type current $J_{N}$ consisting of $N_c$ quark fields
is defined as
\begin{align}
J_{N}(x)
&=\frac{1}{N_{c}!}\varepsilon^{\alpha_{1}\dots\alpha_{N_{c}}}
    \Gamma^{f_{1}\dots f_{N_{c}}}_{(TT_3)(SS_3)}
    \psi_{\alpha_{1}f_{1}}(x)\dots
    \psi_{\alpha_{N_{c}}f_{N_{c}}}(x),
\end{align}
where $\alpha$ and $\beta$ denote color indices, $f$ and
$g$ denote flavor indices, and
$\Gamma$ represents the matrices carrying the flavor and
spin quantum numbers $TT_3$ and $SS_3$ of the nucleon.
As mentioned in Sec.~\ref{sec:2}, the EM form factors of the nucleon
are defined through the matrix element of the EM current given in
Eq.~\eqref{eq:5}. This nucleon matrix element can be expressed as
the following three-point correlation function in Euclidean space:
\begin{align}
\bra{N(p',S')}J_\mu(0)\ket{N(p,S)} &=
  \mathcal{N}^*(p')\mathcal{N}(p)
  \lim _{y_{4}\rightarrow +\infty
\atop x_{4}\rightarrow -\infty } e^{-ip_{4}' y_{4}+ip_{4}x_{4}}
\int d^{3}yd^{3}xe^{-i\bm{p}'\cdot \bm{y}+i\bm{p}\cdot\bm{x}} \cr
&\times  \bra{0} J_{N}(y) (-i)\psi^\dagger \hat{Q} \gamma_\mu \psi
  J^{\dagger }_{N}(x)\ket {0}.
\label{eq:30}
\end{align}
In the large Euclidean-time limit, the three-point correlation function can
be written as
\begin{align}
\bra{0} J_{N}(y) (-i)\psi^\dagger \hat{Q} \gamma_\mu \psi
  J^{\dagger }_{N}(x)\ket {0} &= \frac1{\mathcal{Z}_{\mathrm{eff}}[0]}
  \int \mathcal{D}U  \mathcal{D}\psi^\dagger  \mathcal{D}\psi
 \,\exp\left[
\int d^4 x  \psi_f^\dagger (x) D(U,v) \psi^f (x)
                                \right].
\label{eq:31}
\end{align}
To evaluate the three-point correlation function in
Eq.~\eqref{eq:31}, we introduce the following generating functional in
Euclidean space:
\begin{align}
  \mathcal{W}[\eta^\dagger,\eta,v]
  &=\frac{-i}{\mathcal{Z}_\mathrm{eff}[0]}
  \int \mathcal{D}\psi^\dagger \mathcal{D}\psi \mathcal{D}U
  \exp\left[ \int d^4 x \left\{
    \psi^\dagger (x) (i\rlap{/}{\nabla}+ i\overline{m} \bm{1}) \psi(x)
    -i\eta^\dagger(x) \psi(x) -i\psi^\dagger(x) \eta(x)
    \right. \right. \nonumber \\ & \left.\left.
    +\, i M_0f(\overline{m}) \int
  \frac{d^4k d^4 l}{(2\pi)^8}e^{i(k-l)\cdot x}~
  F(k+\hat{Q}v)F(l+\hat{Q}v)
  \psi^{\dagger}(k) U^{\gamma_5}(x) \psi(l)
 \right\} \right] \cr
&= \frac{-i}{\mathcal{Z}_\mathrm{eff}[0]}
  \int \mathcal{D}U\, \mathrm{Det}\left[D(U,v)\right]
  \exp\left[-\int d^4 x d^4y \,\eta^\dagger (x) \left\langle x\left|
  \frac1{D(U,v)} \right| y \right \rangle \eta(y)\right],
  \label{eq:32}
\end{align}
where $\eta$ denotes the source field for the fermion.
The three-point correlation function in Eq.~\eqref{eq:31} can
then be computed explicitly by taking the functional derivatives of the
generating functional:
\begin{align}
\bra{0} J_{N}(y) (-i)\psi^\dagger(0) \hat{Q} \gamma_\mu \psi (0)
  J^{\dagger }_{N}(x)\ket {0} &= \Gamma^{\{f\}}_{(TT'_3)(SS'_3)}
(\Gamma^{\{g\}}_{(TT_3)(SS_3)})^{*} \frac{\delta}{\delta
     \eta_{f_1}^\dagger (\bm{y}, y_4)}\cdots \frac{\delta}{\delta
     \eta_{f_{N_c}}^\dagger (\bm{y}, y_4)} \frac{\delta}{\delta
  v_\mu(0)} \cr &\times
\left.\mathcal{W}[\eta^\dagger,\eta,v]\right|_{\eta^\dagger,\eta,v=0}
\frac{\overleftarrow{\delta}}{\delta \eta_{g_1} (\bm{x},
  x_4)}\cdots  \frac{\overleftarrow{\delta}}{\delta
     \eta_{g_{N_c}} (\bm{x}, x_4)} .
\end{align}
The result for the EM matrix element can be decomposed into the
valence (level-quark) and sea (Dirac-continuum) contributions:
\begin{align}
  &\mel{N(p',S')}{J_\mu(0)}{N(p,S)}
  = \mel{N(p',S')}{J_\mu(0)}{N(p,S)}_\mathrm{val}
  + \mel{N(p',S')}{J_\mu(0)}{N(p,S)}_{\rm{sea}},
\end{align}
where the valence contribution is given by
\begin{align}
  &\mel{N(p',S')}{J_\mu(0)}{N(p,S)}_\mathrm{val} =
\frac{N_c}{\mathcal{Z}_\mathrm{eff}[0]}
\Gamma^{\{f\}}_{(TT'_3)(SS'_3)}
(\Gamma^{\{g\}}_{(TT_3)(SS_3)})^{*}
\nonumber \\ & \times
  \lim_{T\to\infty}
  e^{-i(p'_{4}+p_{4})\frac{T}{2}}
  \int\;d^{3}xd^{3}y\,
  e^{-i\bm{p}^{\;\prime}\cdot\bm{y}+i\bm{p}\cdot\bm{x}}
  \int \mathcal{D}U\,
  e^{-S_{\text{eff}}[U]}
  \prod^{N_c}_{i=2}\mel**{\bm{y},\frac{T}{2}}{\frac{1}{D[U]}}
  {\bm{x},-\frac{T}{2}}_{f_i,g_i}
  \nonumber \\ & \times
   \mel**{\bm{y},\frac{T}{2}}{\frac{1}{D[U]}}{\bm{0},0}_{f_1,f}
   \Big\{i\gamma_4
    \gamma_\mu \hat{Q}
    - \gamma_4M_0f(\overline{m})
    (\overleftarrow{F}_\mu\hat{Q}U^{\gamma_5}\overrightarrow{F}
    + \overleftarrow{F}U^{\gamma_5}\hat{Q}\overrightarrow{F}_\mu)
   \Big\}_{fg}
   \mel**{\bm{0},0}{\frac{1}{D[U]}}{\bm{x},-\frac{T}{2}}_{g,g_1},
\end{align}
whereas the sea contribution is given by
\begin{align}
  &\mel{N(p',S')}{J_\mu(0)}{N(p,S)}_\mathrm{sea} =
\frac{N_c}{\mathcal{Z}_\mathrm{eff}[0]}
\Gamma^{\{f\}}_{(TT'_3)(SS'_3)}
(\Gamma^{\{g\}}_{(TT_3)(SS_3)})^{*}
\nonumber \\ & \times
  \lim_{T\to\infty}
  e^{-i(p'_{4}+p_{4})\frac{T}{2}}
  \int\;d^{3}xd^{3}y\,
  e^{-i\bm{p}^{\;\prime}\cdot\bm{y}+i\bm{p}\cdot\bm{x}}
  \int \mathcal{D}U\,
  e^{-S_{\text{eff}}[U]}
  \prod^{N_c}_{i=1}\mel**{\bm{y},\frac{T}{2}}
         {\frac{1}{D[U]}}{\bm{x},-\frac{T}{2}}_{f_i,g_i}
  \nonumber \\ & \times
  \tr\bigg[\Big\{i\gamma_4
    \gamma_\mu \hat{Q}
    - \gamma_4M_0f(\overline{m})
    (\overleftarrow{F}_\mu\hat{Q}U^{\gamma_5}\overrightarrow{F}
    + \overleftarrow{F}U^{\gamma_5}\hat{Q}\overrightarrow{F}_\mu)
   \Big\}
  \mel**{\bm{0},0}{\frac{1}{D[U]}}{\bm{0},0}\bigg].
\end{align}
where $F_\mu=\partial F/\partial k_\mu$ is defined as 
the derivative of $F(k)$ with respect to $k_\mu$.
As mentioned above, the path integral over the chiral field $U$
is performed within the saddle-point approximation. However, the
translational and rotational zero modes must be treated exactly.
Explicitly, these zero modes can be incorporated as
\begin{align}
 U_c(\bm{r}) \longrightarrow R(t) T(\bm{Z}(t))
  U_c(\bm{r})T^\dagger(\bm{Z}(t)) R^\dagger(t) = R(t) U(\bm{r} -
  \bm{Z} (t)) R^\dagger (t),
\end{align}
where the rotational matrix $R$ is an element of the SU(2) group.
The translational operator $T(\bm{Z}(t))$ shifts
the position of the mean field to $\bm{r}-\bm{Z}(t)$.
Note that the translational zero modes naturally yield the Fourier
transform of the three-dimensional charge and magnetization
distributions.

The path integration over the $U$ field,
$\mathcal{D}U$, is then replaced by integration over the
zero modes~\cite{Christov:1995vm}
\begin{align}
  \int \mathcal{D}U\,\mathcal{F}[U(x)]\rightarrow
  \int
  \mathcal{D}R\mathcal{D}\bm{Z}\,\mathcal{F}[TRU_c(\bm{r})R^\dagger
  T^\dagger].
\end{align}
The Dirac operator given in Eq.~\eqref{eq:21} can be rewritten as
\begin{align}
  D [U] =  R(t)T(\bm{Z}(t)) i\gamma_4 \left[
  D[U_c] + \frac{i}{2} \Omega^at^a
  \right] T^\dagger(\bm{Z}(t))R^\dagger(t),
\label{eq:39}
\end{align}
where $D[U_c] = \partial_4 + H[U_c]$ and $\Omega^a$ is the angular
velocity defined by
\begin{align}
  R^\dagger\dot{R} = i\Omega = \frac{i}{2}\Omega^a\tau^a.
\end{align}
Note that $t^a$ in Eq.~\eqref{eq:39} contains both local and
nonlocal contributions:
\begin{align}
  t^a = \tau^a + i \gamma_4 M_0 f(\overline{m})\left(
    \overleftarrow{F}_4 \tau^a U^{\gamma_5}_c \overrightarrow{F}
  + \overleftarrow{F}   U^{\gamma_5}_c \tau^a \overrightarrow{F}_4
  \right),
\end{align}
where the second term arises from the momentum-dependent
dynamical quark mass.

Having performed the calculation of the $1/N_c$ rotational
corrections~\cite{Christov:1995hr, Christov:1995vm, Kim:1995mr,
  Choi:2025xha}, we arrive at the final expressions for the EM Sachs
form factors of the nucleon. The isoscalar electric form factor
receives contributions only from the leading order in the $N_c$
expansion:
\begin{align}
  \label{eq:42}
  & G^{T=0}_E(Q^2) = \frac{N_c}{3}\int
    d^3 r\,j_0(|\bm{Q}||\bm{r}|)\mathcal{B}(\bm{r}),
\end{align}
where $\mathcal{B}(\bm{r})$ is identified as the baryonic
density expressed as
\begin{align}
  \mathcal{B}(\bm{r}) &= \mathrm{z_{val}}
  \bra{\mathrm{val}}\ket{\bm{r}}
  \left\{ 1 + i\gamma_4 M_0 f(\overline{m}) \left(
       \overleftarrow{F}_4  U^{\gamma_5}_c \overrightarrow{F}
     + \overleftarrow{F}    U^{\gamma_5}_c \overrightarrow{F}_4
  \right) \right\}
  \bra{\bm{r}}\ket{\mathrm{val}}
   \cr
  &   + \int\frac{d\omega}{2\pi i}\sum_{n}
  \frac{\bra{n_\omega}\ket{\bm{r}}
  \left\{ 1 + i\gamma_4 M_0 f(\overline{m}) \left(
       \overleftarrow{F}_4  U^{\gamma_5}_c \overrightarrow{F}
     + \overleftarrow{F}    U^{\gamma_5}_c \overrightarrow{F}_4
  \right) \right\}
  \bra{\bm{r}}\ket{n_\omega}}
  {\omega + iE_n(\omega)} .
\label{eq:43}
\end{align}
The first term of Eq.~\eqref{eq:43} comes from the valence-quark
contribution, whereas the second one comes from the sea-quark
contribution. To obtain the isovector electric form factor, we have to
take into account the $1/N_c$ rotational corrections:
\begin{align}
  \label{eq:44}
  & G^{T=1}_E(Q^2) = \frac{1}{I} \int d^3 r\,
    j_0(|\bm{Q}||\bm{r}|)\mathcal{I}(\bm{r}),
\end{align}
where $\mathcal{I}(\bm{r})$ denotes the density of the moment of
inertia, given by
\begin{align}
  \mathcal{I}(\bm{r}) &= \frac{N_c}{6}\left(
    \mathrm{z_{val}} \sum_{n}
\frac{\bra{\mathrm{val}}\ket{\bm{r}} t^a \bra{\bm{r}}\ket{n_{\omega_v}}\bra{n_{\omega_v}}
                        t^a \ket{\mathrm{val}}}
    {E_n(\omega_v)-E_\mathrm{val}}
    + \frac12 \mathrm{z_{val}} \bra{\mathrm{val}}\ket{\bm{r}} T^{aa}
                        \bra{\bm{r}}\ket{\mathrm{val}}
    \right. \nonumber \\ & \left.
    + \frac12 \int \frac{d\omega}{2\pi}  \sum_{n,m}
    \frac{\bra{m_\omega}\ket{\bm{r}} t^a
                           \bra{\bm{r}}\ket{n_\omega}\bra{n_\omega}
                           t^a \ket{m_\omega}}
    {[\omega+iE_n(\omega)][\omega+iE_m(\omega)]}
    + \frac12 \int \frac{d\omega}{2\pi i}\sum_{n}
    \frac{\bra{n_\omega}\ket{\bm{r}} T^{aa} \bra{\bm{r}}\ket{n_\omega}}
    {\omega + iE_n(\omega)}
                           \right).
\label{eq:45}
\end{align}
where $\omega_v=-iE_{\mathrm{val}}(\omega_v)$. A detailed
discussion for this quantity is in Ref.~\cite{Broniowski:2001cx}.
The operators $t^a$ and $T^{ab}$ in isospin space in Eq.~\eqref{eq:45}
are defined as
\begin{align}
  t^a &= \tau^a + i \gamma_4 M_0 f(\overline{m})\left(
    \overleftarrow{F}_4 \tau^a U^{\gamma_5}_c \overrightarrow{F}
  + \overleftarrow{F}   U^{\gamma_5}_c \tau^a \overrightarrow{F}_4
  \right), \\
  T^{ab} &= \gamma_4 M_0 f(\overline{m}) \left(
   \overleftarrow{F}_{44} \tau^a \tau^b U^{\gamma_5}_c
   \overrightarrow{F}
   + 2 \overleftarrow{F}_4    \tau^a
   U^{\gamma_5}_c \tau^b
   \overrightarrow{F}_4
   +   \overleftarrow{F}
   U^{\gamma_5}_c \tau^a \tau^b \overrightarrow{F}_{4 4}\right).
\end{align}
The valence and sea contributions to the moment of inertia are given
by\footnote{The expressions for the moment of inertia in
  Ref.~\cite{Choi:2025xha} contain typographical errors, which are
  corrected in the present work.}
\begin{align}
I_{\mathrm{val}}\delta^{ab} &=
\frac{N_{c}}{2}\mathrm{z}_{\mathrm{val}}\sum_{n}\frac{
  \langle \mathrm{val}| t^{a} | n_{\omega_v}\rangle 
  \langle n_{\omega_v}| t^{b}| \mathrm{val}\rangle
}{E_{n}(\omega_v)-E_{\mathrm{val}}}
+ \frac{N_{c}}{4} \mathrm{z}_{\mathrm{val}}\langle
\mathrm{val}  | T^{ab} | \mathrm{val} \rangle,\cr
I_{\mathrm{sea}}\delta^{ab} &= \frac{N_{c}}{4} \sum_{n,m}
\int \frac{d\omega}{2\pi}~  \frac{\langle m_\omega| t^{a}|
n_\omega\rangle}{\omega+i E_{n}(\omega)}  \frac{\langle n_\omega| 
t^{b}| m_\omega\rangle}{\omega+i E_{m}(\omega)} + \frac{N_{c}}{4}
\sum_{n}  \int\frac{d\omega}{2\pi i}
\frac{\langle n_\omega| T^{ab}| n_\omega\rangle}{
\omega+i E_{n}}.
\label{eq:48}
\end{align}
If we switch off the momentum dependence of the dynamical quark mass,
the results for the electric form factors of the nucleon reduce
to those in the $\chi$QSM~\cite{Christov:1995hr, Christov:1995vm}
without regularization.

The isoscalar magnetic form factor again arises from the
next-to-leading-order contribution in the $1/N_c$ expansion:
\begin{align}
   G^{T=0}_M(Q^2) &= \frac{N_c M_N}{4I} \int\, d^3 r
      \frac{j_1(|\bm{Q}||\bm{r}|)}{|\bm{Q}||\bm{r}|}
  \mathcal{X}(\bm{r}),
  \label{eq:49}
\end{align}
where the density $\mathcal{X}(\bm{r})$ is given by
\begin{align}
  \mathcal{X}(\bm{r}) &=
  2\mathrm{z_{val}} \sum_{n}
  \frac{\bra{\mathrm{val}}\ket{\bm{r}} t^z_{M,T=0}
        \bra{\bm{r}}\ket{n_\omega}
        \bra{n_\omega}   t^z   \ket{\mathrm{val}}}
       {E_n(\omega)-E_\mathrm{val}}
  + \mathrm{z_{val}} \bra{\mathrm{val}}\ket{\bm{r}} iT^{zz}_{M,T=0}
                        \bra{\bm{r}}\ket{\mathrm{val}}
  \nonumber \\ &
  + \int \frac{d\omega}{2\pi} \sum_{n,m}
  \frac{\bra{m_\omega}\ket{\bm{r}}t^z_{M,T=0}\bra{\bm{r}}\ket{n_\omega}
        \bra{n_\omega}    t^z   \ket{m_\omega}}
       {[\omega+iE_n(\omega)][\omega+iE_m(\omega)]}
  + \int \frac{d\omega}{2\pi i}\sum_{n}
  \frac{\bra{n_\omega}\ket{\bm{r}} iT^{zz}_{M,T=0} \bra{\bm{r}}\ket{n_\omega}}
       {\omega + iE_n(\omega)} .
\end{align}
$t^i_{M,T=0}$ and $T^{ia}_{M,T=0}$ are defined respectively as
\begin{align}
  t^i_{M,T=0} = & -\gamma_5(\bm{r}\times\bm{\sigma})_i
  - \gamma_4 M_0 f(\overline{m}) \left(
    {\overleftarrow{F}'} L_iU^{\gamma_5}_c
                  \overrightarrow{F}
  +        \overleftarrow{F}       U^{\gamma_5}_c L_i
                  \overrightarrow{F}' \right)
\end{align}
and
\begin{align}
  T^{ia}_{M,T=0} =
  \gamma_4 M_0 f(\overline{m}) \left(
      {\overleftarrow{F}'}_{4} L_i \tau^a U^{\gamma_5}_c
  \overrightarrow{F}
    + {\overleftarrow{F}'} L_i U^{\gamma_5}_c \tau^a
  \overrightarrow{F} _{4}
    +        \overleftarrow{F} _{4} \tau^a U^{\gamma_5}_c L_i
  {\overrightarrow{F}'}
    +        \overleftarrow{F}       U^{\gamma_5}_c \tau^a L_i
  {\overrightarrow{F}'}_{4}    \right),
\end{align}
where $F'=\partial F/\partial k^2$ is defined as 
the derivative of $F(k)$ with respect to $k^2$ and $L_i$ is the orbital angular
momentum operator.
The isovector magnetic form factor contains both the
leading-order and $1/N_c$-order contributions:
\begin{align}
  & G^{T=1}_M(Q^2) = N_c
    M_N\int\,d^3\bm{r}\frac{j_1(|\bm{Q}||\bm{r}|)}{|\bm{Q}||\bm{r}|}
  \left( \mathcal{Q}_0(\bm{r}) - \frac{1}{4I} \mathcal{Q}_1(\bm{r})
    \right),
\label{eq:52}
\end{align}
where $\mathcal{Q}_0(\bm{r})$ and $\mathcal{Q}_1(\bm{r})$ are given by
\begin{align}
  \mathcal{Q}_0(\bm{r}) =
  \mathrm{z_{val}} \bra{\mathrm{val}}\ket{\bm{r}} t^{zz}_{M,T=1}
  \bra{\bm{r}}\ket{\mathrm{val}}
  + \int\frac{d\omega}{2\pi i}\sum_{n}
  \frac{\bra{n_\omega}\ket{\bm{r}} t^{zz}_{M,T=1}
  \bra{\bm{r}}\ket{n_\omega}}{\omega + iE_n(\omega)}
\end{align}
and
\begin{align}
  \mathcal{Q}_1(\bm{r}) &= i\epsilon_{z ij} \Bigg[
    2\mathrm{z_{val}} 
      \sum_{n}\mathrm{z}_n\mathrm{sign}(E_n)
      \frac{\bra{\mathrm{val}}\ket{\bm{r}} t^{zj}_{M,T=1}
                          \bra{\bm{r}}\ket{n_{\omega_n}}
            \bra{n_{\omega_n}} t^i \ket{\mathrm{val}}}
           {E_n(\omega_n)-E_\mathrm{val}(\omega_v)}
    + \mathrm{z_{val}} \bra{\mathrm{val}}\ket{\bm{r}} iT^{zij}_{M,T=1}
      \bra{\bm{r}}\ket{\mathrm{val}}
  \nonumber \\ &
    + \int\frac{d\omega_1}{2\pi}\frac{d\omega_2}{2\pi}\,
    \mathcal{P}\left( \frac{2i}{\omega_2-\omega_1} \right)
    \sum_{n,m}
    \frac{\bra{m_{\omega_1}}\ket{\bm{r}}\,t^{zj}_{M,T=1}
                 \bra{\bm{r}}\ket{n_{\omega_2}}
                 \bra{n_{\omega_2}} t^i \ket{m_{\omega_1}}}
    {[\omega_1+iE_m(\omega_1)][\omega_2+iE_n(\omega_2)]}
  \nonumber \\ &
    + \int\frac{d\omega}{2\pi i}\sum_{n_\omega}
    \frac{\bra{n_\omega}\ket{\bm{r}} iT^{zij}_{M,T=1}
                 \bra{\bm{r}}\ket{n_\omega}
    }{\omega + iE_n(\omega)}
  \Bigg],
\end{align}
where $\omega_n = -iE_n(\omega_n)$ and $\mathcal{P}$ stands for
the Cauchy principal value.
The operators $t^{ia}_{M,T=1}$ and $T^{iab}_{M,T=1}$ are defined respectively as
\begin{align}
  t^{ia}_{M,T=1}
  &= -\gamma_5 (\bm{r}\times\bm{\sigma})_i \tau^a
  - \gamma_4 M_0 f(\overline{m}) \left(
   {\overleftarrow{F}'} L_i  \tau^a U^{\gamma_5}_c
    \overrightarrow{F}
  +       \overleftarrow{F}      U^{\gamma_5}_c \tau^a L_i
    {\overrightarrow{F}'} \right)
\end{align}
and
\begin{align}
  T^{iab}_{M,T=1}
  = \gamma_4 M_0 f(\overline{m}) \left(
  -{\overleftarrow{F}'}_{4} L_i \tau^a \tau^b U^{\gamma_5}_c
  \overrightarrow{F}
  + {\overleftarrow{F}'}   L_i \tau^b U^{\gamma_5}_c \tau^a
  \overrightarrow{F}_4
  -        \overleftarrow{F}_4    \tau^a U^{\gamma_5}_c \tau^b L_i
  {\overrightarrow{F}'}
  +        \overleftarrow{F}      U^{\gamma_5}_c \tau^b \tau^a L_i
  {\overrightarrow{F}'}_{4}
  \right).
\end{align}
Once again, the results for the magnetic form factors of the nucleon become
identical to the nonregularized expressions of the
$\chi$QSM~\cite{Christov:1995hr, Christov:1995vm} when the momentum
dependence of the dynamical quark mass is switched off.

\section{Results and discussion}
\label{sec:4}
Since we start from the improved effective low-energy QCD partition
function given in Eq.~\eqref{eq:15} beyond the chiral limit, the
average size of the instanton and the average interdistance between
instantons are fixed to $\bar{\rho}=0.35$ fm and $\bar{R}=0.86$ fm,
respectively. In addition, we take the average value of the current
up- and down-quark masses to be $\overline{m}=5$
MeV~\cite{Goeke:2007bj}, which yields the pion mass
$m_\pi=140$ MeV. The dynamical quark mass at zero virtuality is
determined to be $M_0=385$ MeV from the saddle-point equation beyond
the chiral limit~\cite{Goeke:2007bj, Goeke:2007nc}. Thus, we have no
adjustable free parameter in the present calculation.
In Ref.~\cite{Choi:2025xha}, the effective chiral theory based on the
instanton vacuum was developed in the chiral limit. Therefore, we need
to compute the classical nucleon mass and the moment of inertia with
the current quark mass taken into account. The momentum-dependent
quark mass in the valence-quark part gives rise to fundamental
complications related to the analytic continuation from Euclidean to
Minkowski space, since the momentum-dependent dynamical quark mass
derived from the instanton vacuum is strictly defined in Euclidean
space. For this reason, we retain the momentum dependence of the
dynamical quark mass only in the sea-quark (Dirac-continuum) part.
This approximation is nevertheless justified, since the valence part
is finite and thus does not require any regularization. For the same
reason, the valence part was also not regularized in the $\chi$QSM.
Finally, we want to mention that the momentum-dependent dynamical
quark mass requires a much larger box size than that in the $\chi$QSM
in order to ensure numerical stability. We use a box size of
$D_{\mathrm{box}}\approx 11$ fm. Note that we solve the
single-particle Dirac equation by using the Kahana-Ripka
method~\cite{Kahana:1984be}.

\begin{table}[htp]
  \centering
    \caption[The classical nucleon mass for each case]
  {Classical nucleon mass $M_{\text{cl}}$. The second
    row presents the results of the present work, whereas the third row
    corresponds to those obtained in the chiral limit. The fourth and
    fifth rows show the results obtained in the $\chi$QSM with the dynamical
    quark mass $M_0=385$ MeV and $M_0=420$ MeV, respectively.}
  \label{tab:1}
  \renewcommand{\arraystretch}{2}
  \setlength{\tabcolsep}{9pt}
  \begin{tabular}{ c c c c c }
  \hline\hline
                                &  $E_{\text{val}}$ [MeV]  &  $N_cE_{\text{val}}$ [MeV]  &   $E_{\text{sea}}$ [MeV]   &   $M_{\text{cl}}$ [MeV]  \\ \hline
  This work   ($M_0$=$385$ MeV) &           162.2          &              486.5          &            530.1           &          1017            \\
  Chiral limit($M_0$=$359$ MeV) &           152.6          &              457.7          &            529.7           &          987.4           \\
     $\chi$QSM($M_0$=$385$ MeV) &           223.3          &              669.8          &            588.0           &          1258            \\
     $\chi$QSM($M_0$=$420$ MeV) &           203.5          &              610.6          &            645.8           &          1256            \\
  \hline\hline
  \end{tabular}
\end{table}
The classical nucleon mass~\cite{Diakonov:1987ty} is expressed as
\begin{align}
    M_{\mathrm{cl}} &=\underset{\pi(x)}{\mathrm{min}} \left[N_c
               E_{\mathrm{val}} + E_{\mathrm{sea}} \right],
\end{align}
where $E_{\mathrm{val}}$ and $E_{\mathrm{sea}}$ stand for the
valence-quark and sea-quark energies, respectively. The saddle-point
approximation, which we apply to the integration over the chiral
field, gives rise to the equation of motion for the classical pion
field. Its solution defines the pion mean field and is determined
through a self-consistent iterative procedure. Details of the
minimization scheme for $M_{\mathrm{cl}}$ can be found in
Refs.~\cite{Christov:1995vm, Choi:2025xha}. Table~\ref{tab:1}
summarizes the present numerical results for the classical nucleon
mass together with those obtained within the $\chi$QSM. Since the
current quark mass is incorporated here, we further compare
our values with the corresponding ones in the chiral limit. For the
chiral-limit case, we follow Ref.~\cite{Diakonov:1985eg}, in which the
instanton parameters are fixed to $\bar{\rho}=1/3$ fm and $\bar{R}=1$
fm, while the dynamical quark mass at zero virtuality is determined to
be $M_0=359$ MeV. As seen by comparing the second and third rows, the
current quark mass mainly influences the valence-quark part, leaving
the sea-quark part nearly unchanged.

In the fourth and fifth rows, we list the results from the $\chi$QSM
with the values of the dynamical quark mass $M_0=385$ MeV and 420 MeV,
respectively. In the $\chi$QSM, $M_0=420$ MeV is selected, because it
yields the best results for baryonic observables in comparison with
experimental data. However, the constant $M$ inevitably requires a
regularization scheme to tame quark loops. For the present comparison,
we adopt the proper-time regularization, in which the cutoff mass
$\Lambda$ is fixed so as to reproduce the experimental value of the
pion decay constant $f_\pi$. A single cutoff, however, leads to a
current quark mass of $m_{\mathrm{u}}=m_{\mathrm{d}}\approx 15$ MeV
when the pion mass is constrained to $m_\pi=140$
MeV~\cite{Blotz:1992pw}. Compared with the results from the $\chi$QSM,
the present ones are consistently smaller. As a result, the value of
the classical nucleon mass turns out to be closer to the experimental
value of the nucleon mass.

\begin{table}[htp]
  \centering
    \caption[The moment of inertia for each cases]
  {Moments of inertia in units of fm. The second
    row presents the results of the present work, whereas the third row
    corresponds to those obtained in the chiral limit. The fourth and
    fifth rows show the results obtained in the $\chi$QSM with the
    dynamical quark mass $M_0=385$ MeV and $M_0=420$ MeV,
    respectively. In the last column, the results for the
    $\Delta$--$N$ mass splitting are given in units of MeV.}
  \label{tab:2}
\renewcommand{\arraystretch}{2}
  \setlength{\tabcolsep}{9pt}
  \begin{tabular}{ c c c c c }
  \hline\hline
                                & $I_{\text{val}}$ [fm]  & $I_{\text{sea}}$ [fm]  &  $I$ [fm]  &   $M_{\Delta-N}$ [MeV]  \\ \hline
  This work   ($M_0$=$385$ MeV) &          1.075         &           0.089        &   1.164    &          254.2          \\
  Chiral limit($M_0$=$359$ MeV) &          1.085         &           0.315        &   1.400    &          211.5          \\
     $\chi$QSM($M_0$=$385$ MeV) &          0.949         &           0.248        &   1.197    &          247.3          \\
     $\chi$QSM($M_0$=$420$ MeV) &          0.775         &           0.255        &   1.031    &          287.2          \\
  \hline\hline
  \end{tabular}
\end{table}
In Table~\ref{tab:2}, we list the numerical results for the moment of  
inertia, which is defined in Eq.~\eqref{eq:48}, which are also
compared with those from the $\chi$QSM. Comparing the results in the
second row with those in the third row, we observe that the
introduction of the current quark mass suppresses the sea-quark
contribution to the moment of inertia. The value of $I$ directly
determines the $\Delta -N$ mass difference ~\cite{Adkins:1983ya,
Choi:2025xha, Diakonov:1987ty}, which is given by 
\begin{align}
M_{\Delta -N} = M_\Delta - M_N = \frac{3}{2I}  .
\end{align}
This work yields $M_{\Delta -N}=254.2$ MeV, which is closer to the
experimental center value of the $\Delta$ isobar, 
$(M_{\Delta}-M_N)_{\mathrm{Exp}}\approx (270-290)$
MeV~\cite{ParticleDataGroup:2024cfk}.   

\begin{figure}[htp]
  \centering
  \includegraphics[scale=0.18]{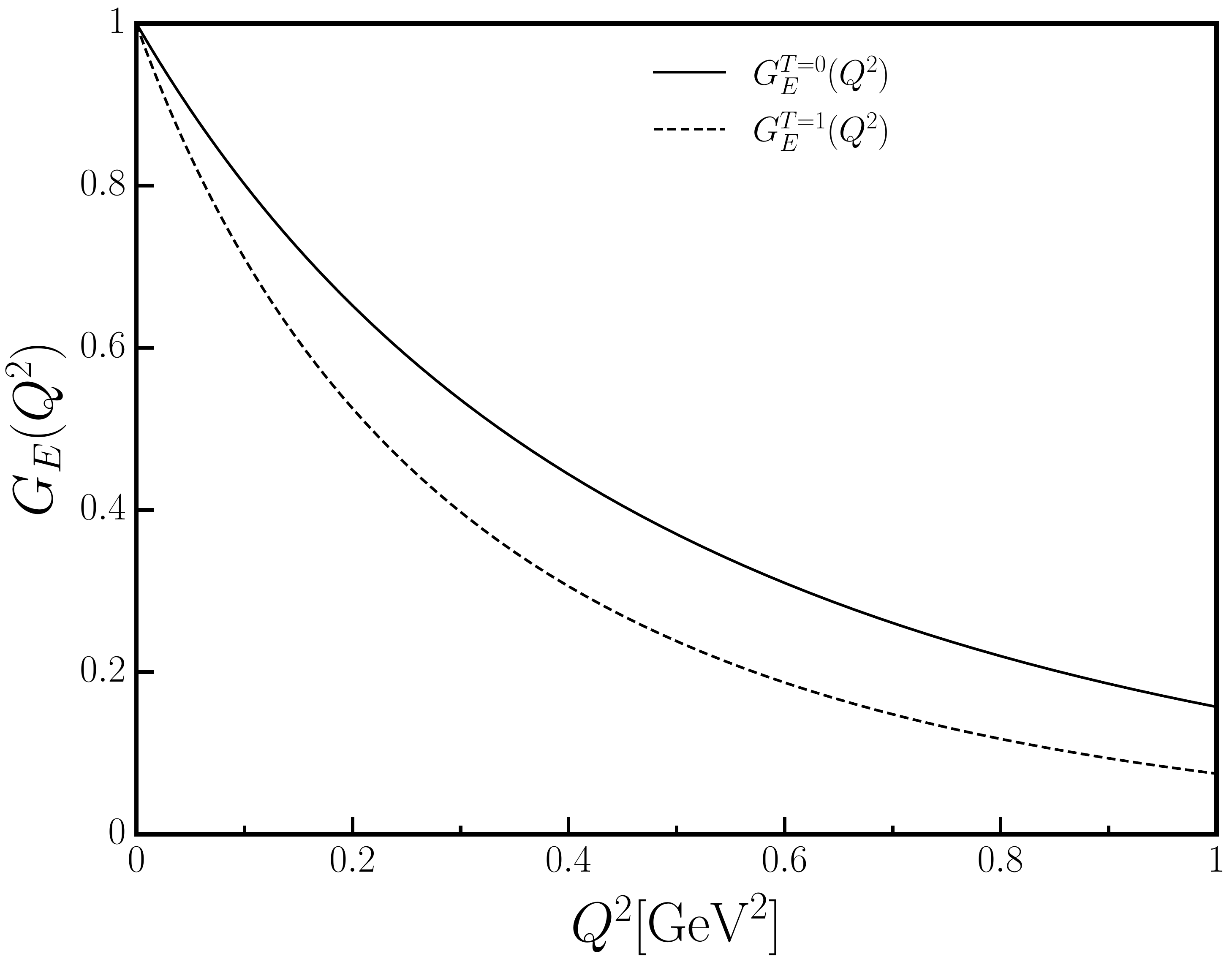}
  \caption[Isoscalar and isovector electric form factors.]
  {Isoscalar and isovector electric form factors of the nucleon. The
    solid curve represents the result for the isoscalar electric form
    factor, whereas the dashed one denotes that for the isovector
    electric form factor.}
\label{fig:1}
\end{figure}
In Fig.~\ref{fig:1}, the numerical results for the isoscalar and
isovector electric form factors are shown. As given in
Eq.~\eqref{eq:42}, $G_E^{T=0}(Q^2)$ arises from the baryon density,
whereas $G_E^{T=1}(Q^2)$, given in Eq.~\eqref{eq:44}, originates from
the $1/N_c$ rotational corrections. Thus, the contribution from the
baryon density is of leading order in the large $N_c$ expansion. As
expected from this $N_c$ counting, the isoscalar electric form factor
is larger than the isovector one over the entire $Q^2$ range
considered.

\begin{figure}[h]
  \centering
  \includegraphics[scale=0.18]{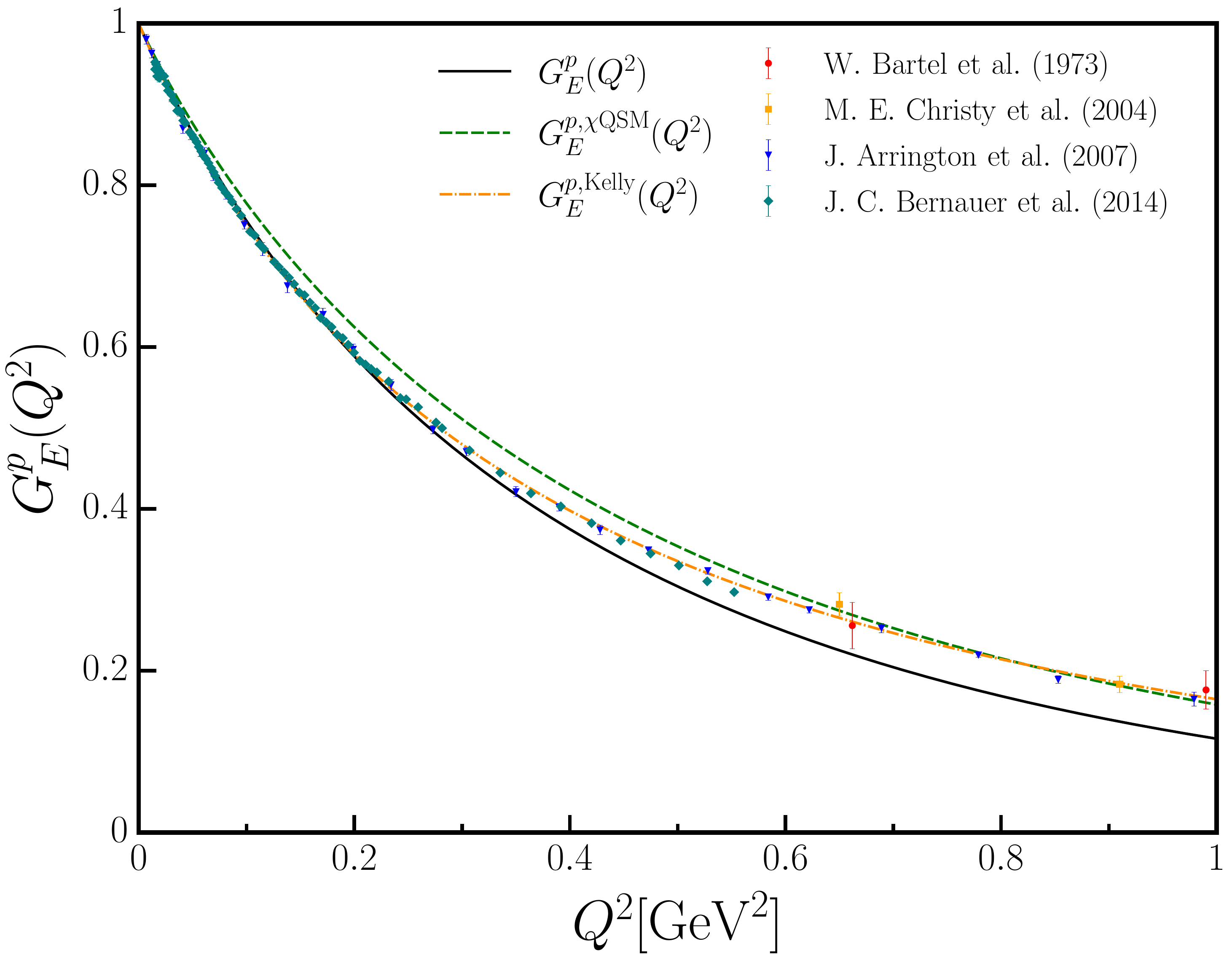}
  \caption[Electric form factor of the proton.]
  {Electric form factor of the proton. The solid curve depicts the
    present numerical result for the proton electric form factor,
    while the dashed and dot-dashed ones represent those obtained
    from the $\chi$QSM with $M_0=420$
    MeV~\cite{Christov:1995hr,Christov:1995vm} and from the Kelly
    parametrization~\cite{Kelly:2004hm}, respectively. The
    experimental data are taken from
    Refs.~\cite{Bartel:1973rf,E94110:2004lsx,Arrington:2007ux,A1:2013fsc}.
  }
\label{fig:2}
\end{figure}
\begin{figure}[htp]
  \centering
  \includegraphics[scale=0.18]{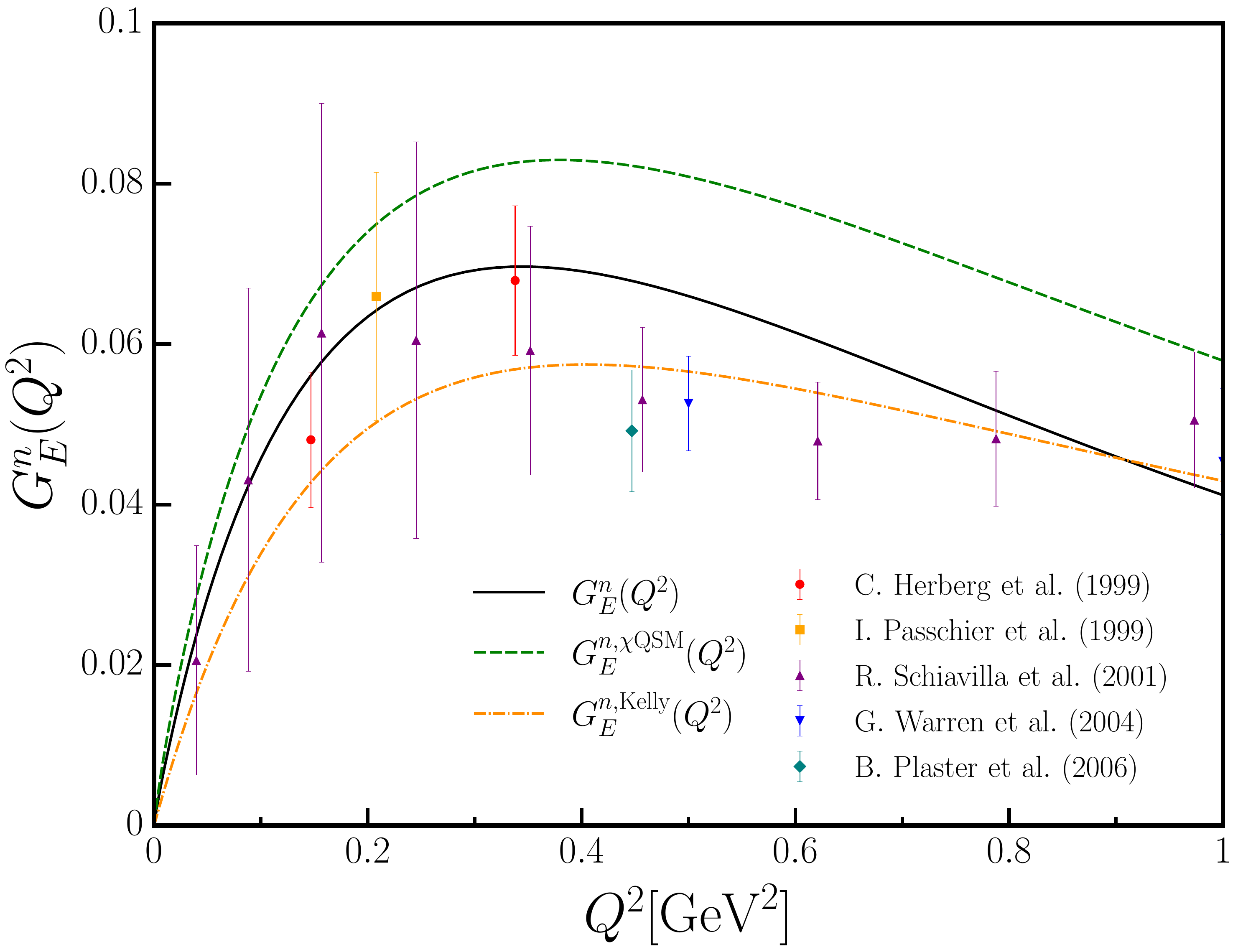}
  \caption[Electric form factor of the neutron.]
  {Electric form factor of the neutron. The solid curve depicts the
    present numerical result for the neutron electric form factor,
    while the dashed and dot-dashed ones represent those obtained
    from the $\chi$QSM with $M_0=420$
    MeV~\cite{Christov:1995hr,Christov:1995vm} and from the
    Galster-type parametrization given in
    Ref.~\cite{Kelly:2004hm}, respectively. The experimental data
    are taken from Refs.~\cite{Herberg:1999ud,Passchier:1999cj,
      Schiavilla:2001qe, JeffersonLabE93-026:2003tty,
      JeffersonLaboratoryE93-038:2005ryd}.}
\label{fig:3}
\end{figure}
In Fig.~\ref{fig:2}, we draw the numerical result for the proton
electric form factor. The dashed and dot-dashed curves depict those
from the $\chi$QSM and from the Kelly
parametrization~\cite{Kelly:2004hm}, respectively. The present result
is in remarkable agreement with the experimental data in the lower
$Q^2$ region ($Q^2 \le 0.4\,\mathrm{GeV}^2$). As $Q^2$ increases,
however, it falls off faster than the data. While the sea-quark
contribution amounts to only about $5\%$, it is essential for
satisfying the correct normalization of the proton charge. The
contribution from the nonlocal part of the EM current is negligible
for the proton electric form factor. Figure~\ref{fig:3} presents the
numerical result for the neutron electric form factor, which is in
very good agreement with the experimental data. The present result
describes the data noticeably better than the $\chi$QSM, which tends
to overestimate $G_E^n(Q^2)$ in the intermediate $Q^2$ region. In
comparison with the Galster-type parametrization given in
Ref.~\cite{Kelly:2004hm}, the present work provides a better
description in the lower $Q^2$ region ($Q^2 \le 0.4\,\mathrm{GeV}^2$),
whereas the latter follows the data more closely as $Q^2$ increases.
Nevertheless, both reproduce the overall $Q^2$ dependence of the
data.

%
\begin{table}[htp]
  \centering
  \caption[Charge radii of the nucleon]{Proton charge radius and
  neutron mean-square charge radius.}
\renewcommand{\arraystretch}{2}
\setlength{\tabcolsep}{9pt}
\begin{tabular}{ c c c c }
\hline \hline
         & \phantom{-} This work & \phantom{-} $\chi$QSM & \phantom{$-$}Experiments~\cite{ParticleDataGroup:2024cfk} \\ \hline
 $ \sqrt{\langle r^2 \rangle_{\mathrm{ch}}^p} $ [$\mathrm{fm}  $] & \phantom{$-$}0.841
                                 &  \phantom{$-$}0.815   &
  $\phantom{-}  0.8409\pm 0.0004$  \\ 
 $       \langle r^2 \rangle_{\mathrm{ch}}^n  $ [$\mathrm{fm}^2$] &
    $-$0.168   & $-$0.219  & $ -0.1155\pm
  0.0017$
  \\ 
\hline \hline
\end{tabular}
\label{tab:3}
\end{table}
In Table~\ref{tab:3}, we list the results for the proton charge
radius and the neutron mean-square charge radius. The present result
for the proton charge radius is in remarkable agreement with the
experimental data, while that for the neutron mean-square charge
radius shows reasonable agreement. Both results are in better
agreement with experiment than those obtained from the $\chi$QSM.

\begin{figure}[htp]
  \centering
  \includegraphics[scale=0.18]{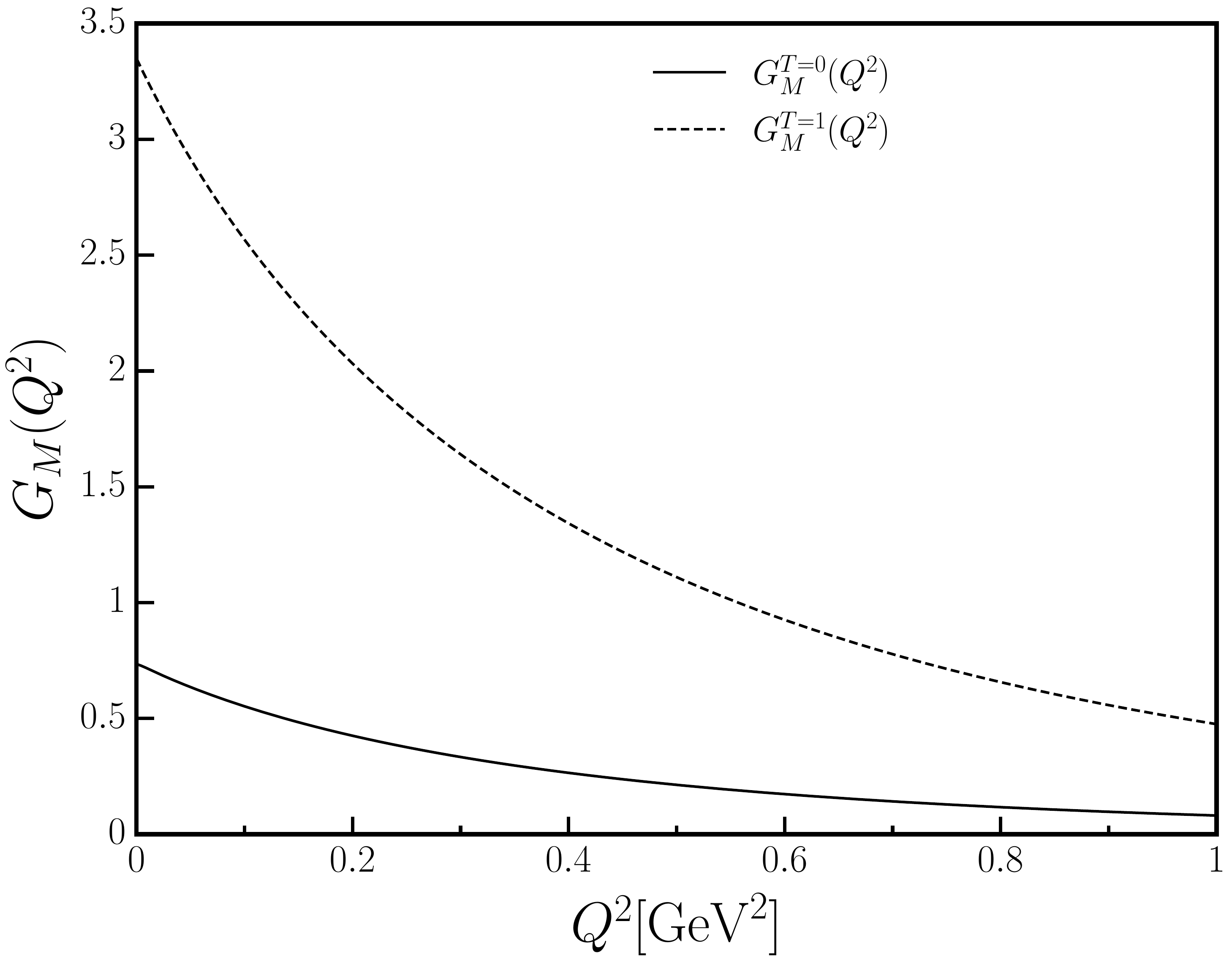}
  \caption[Isoscalar and isovector magnetic form factors of the nucleon]
  {Isoscalar and isovector magnetic form factors of the nucleon. The
    solid curve represents the result for the isoscalar magnetic form
    factor, whereas the dashed one denotes that for the isovector
    magnetic form factor.}
  \label{fig:4}
\end{figure}
Figure~\ref{fig:4} shows the results for the isoscalar and isovector
magnetic form factors of the nucleon, which are defined in
Eqs.~\eqref{eq:49} and~\eqref{eq:52}, respectively. The isoscalar
magnetic form factor arises solely from the $1/N_c$ rotational
corrections, whereas the isovector one contains both the leading and
$1/N_c$ rotational contributions. As expected from this $N_c$
counting, the isovector magnetic form factor dominates over the
isoscalar one, as clearly seen in Fig.~\ref{fig:4}.
\begin{figure}[htp]
  \centering
  \includegraphics[scale=0.18]{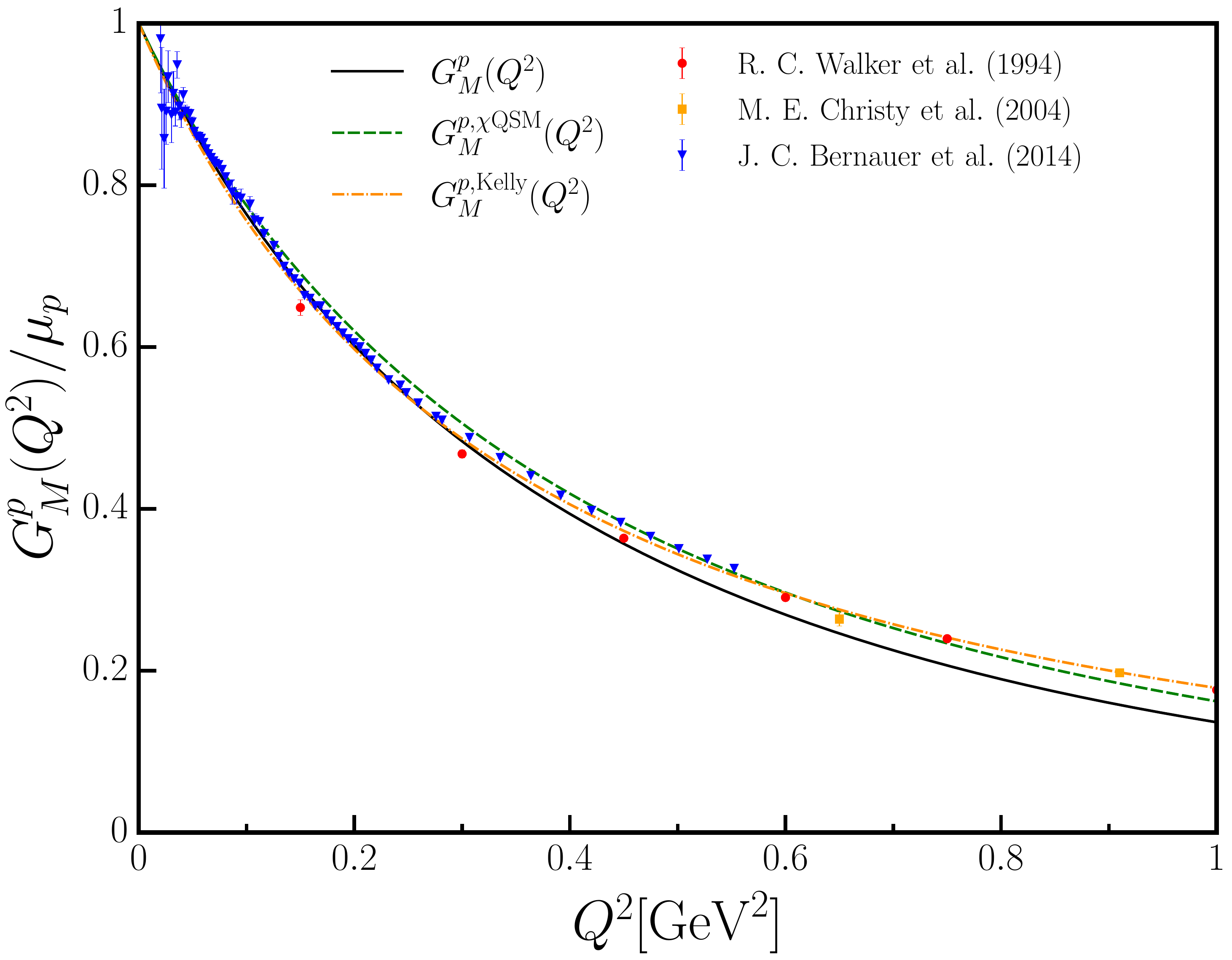}
  \caption[Magnetic form factor of the proton]
  {Magnetic form factor of the proton normalized
  by its magnetic moment. The solid curve depicts the
    present numerical result for the proton magnetic form factor,
    while the dashed and dot-dashed ones represent those obtained
    from the $\chi$QSM with $M_0=420$
    MeV~\cite{Christov:1995hr,Christov:1995vm} and from the Kelly
    parametrization~\cite{Kelly:2004hm}, respectively. The
    experimental data are taken from
    Refs.~\cite{E94110:2004lsx,A1:2013fsc,Walker:1993vj}.}
  \label{fig:5}
\end{figure}
\begin{figure}[htp]
  \centering
  \includegraphics[scale=0.18]{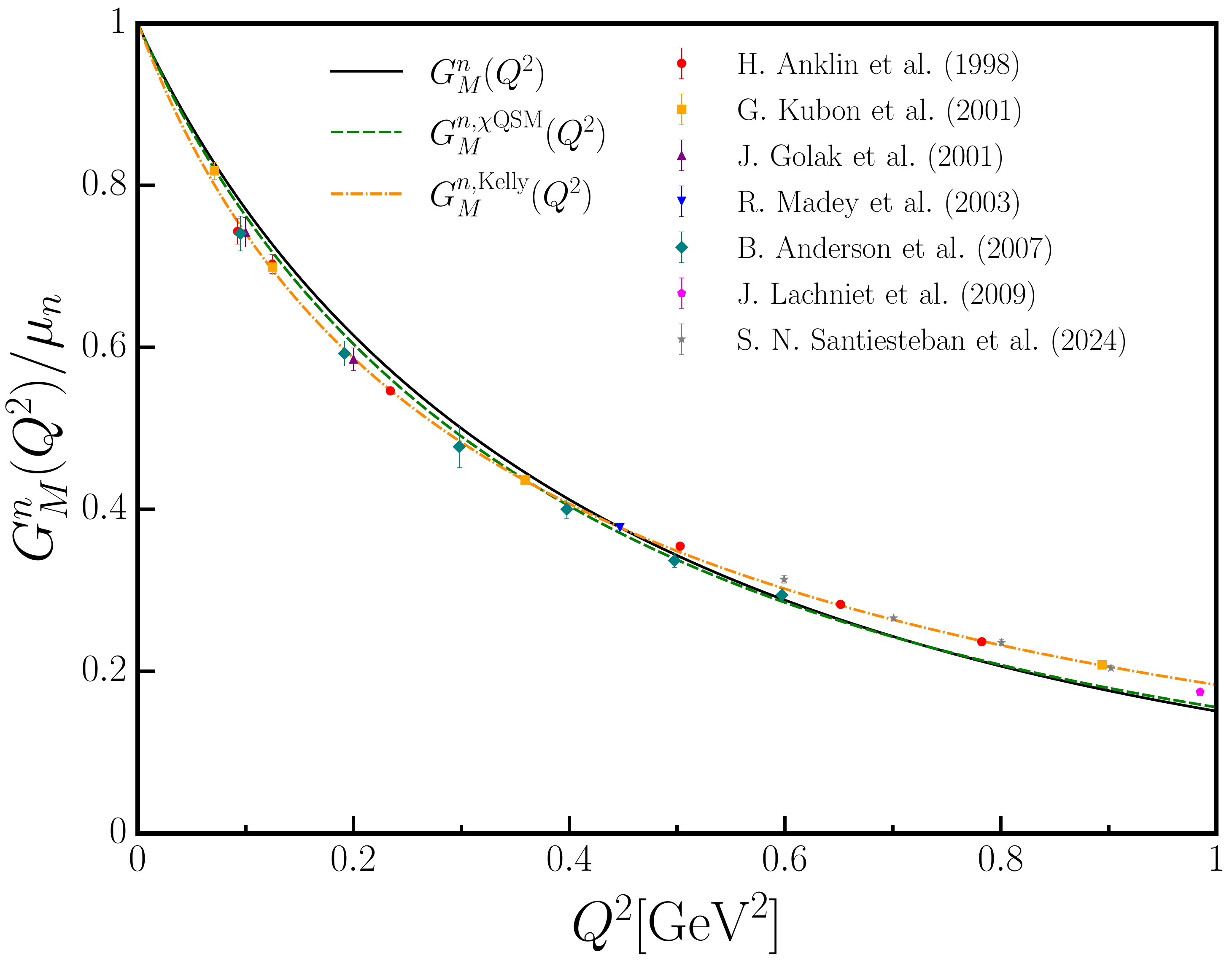}
  \caption[Magnetic form factor of the neutron]
  {Magnetic form factor of the neutron normalized
  by its magnetic moment. The solid curve depicts the
    present numerical result for the neutron magnetic form factor,
    while the dashed and dot-dashed ones represent those obtained
    from the $\chi$QSM with $M_0=420$
    MeV~\cite{Christov:1995hr,Christov:1995vm} and from the Kelly
    parametrization~\cite{Kelly:2004hm}, respectively. The
    experimental data are taken from
    Refs.~\cite{Anklin:1998ae,Kubon:2001rj,Golak:2000nt,
      E93-038:2003ixb,JeffersonLabE95-001:2006dax,CLAS:2008idi,
      JeffersonLabHallA:2023rsh}.} 
  \label{fig:6}
\end{figure}
In Fig.~\ref{fig:5}, we present the results for the proton magnetic
form factor normalized by its magnetic moment. As in the case of the
proton electric form factor, the present result describes the
experimental data very well in the lower $Q^2$ region ($Q^2 \lesssim
0.4\,\mathrm{GeV}^2$). As $Q^2$ increases, however, it falls off
faster than the data. Figure~\ref{fig:6} shows the results for the
normalized magnetic form factor of the neutron, which turns out to be
very similar to the corresponding result from the $\chi$QSM. The
$Q^2$ dependence of the neutron magnetic form factor deviates
slightly from the data. 

%
\begin{table}[htp]
  \centering
  \caption[Magnetic moments of isoscalar, 
isovector and the nucleon]{Magnetic moments 
of the proton and neutron given in units of the nuclear 
magneton $\mu_N$. The results from the $\chi$QSM are evaluated with
$M_0=420$ MeV. The experimental values are taken from
Ref.~\cite{ParticleDataGroup:2024cfk},   
which are rounded to three decimal places.}
\renewcommand{\arraystretch}{2}
\begin{tabular}{c c | c c c}
\hline
\hline
\multicolumn{1}{c|}{Quantities }& types   & This work  & $\chi$QSM &
  \phantom{--}Experiments~\cite{ParticleDataGroup:2024cfk} \\[1.5pt] \hline 
\multicolumn{1}{c|}{           }& valence & $  1.724 $ & $  1.564 $ & --                                \\[1.5pt]
\multicolumn{1}{c|}{ $\mu_{p}$ }& sea     & $  0.312 $ & $  0.378 $ & --                                \\[1.5pt]
\multicolumn{1}{c|}{           }& total   & $  2.042 $ & $  1.942 $ &
  $ 2.793 $   \\[1.5pt] \hline 
\multicolumn{1}{c|}{           }& valence & $ -1.015 $ & $ -0.860 $ & --                                \\[1.5pt]
\multicolumn{1}{c|}{ $\mu_{n}$ }& sea     & $ -0.292 $ & $ -0.455 $ & --                                \\[1.5pt]
\multicolumn{1}{c|}{           }& total   & $ -1.307 $ & $ -1.315 $ &
    $-1.913 $   \\[1.5pt]
\hline
\hline
\end{tabular}

\label{tab:4}
\end{table}
\begin{table}[htp]
  \centering
  \caption[Magnetization radii of the nucleon]
{Magnetization radii of the proton and neutron. 
The results from the $\chi$QSM are evaluated with $M_0=420$ MeV.}
\renewcommand{\arraystretch}{2}
\setlength{\tabcolsep}{9pt}
\begin{tabular}{ c c c c }
\hline \hline
 & This work & $\chi$QSM & \phantom{$-$}Experiments~\cite{ParticleDataGroup:2024cfk} \\ \hline
$\sqrt{\langle r^2 \rangle_M^p}$ [$\mathrm{fm}$] & \phantom{$-$}0.832 & \phantom{$-$}0.810 & \phantom{$-$}0.851$\pm$0.026 \\
$\sqrt{\langle r^2 \rangle_M^n}$ [$\mathrm{fm}$] &
                                                     \phantom{$-$}0.817
             & \phantom{$-$}0.844 & \phantom{$-$}
                                    $0.864_{-0.008}^{+0.009} $\\ 
\hline \hline
\end{tabular}

\label{tab:5}
\end{table}
In Table~\ref{tab:4}, we list the results for the magnetic moments of
the proton and neutron in units of the nuclear magneton.  
It is well known that the magnetic moments of nucleons are
underestimated almost by 30\% in both the SU(2)~\cite{Christov:1995hr}  
and SU(3)~\cite{Kim:1995mr} $\chi$QSM. Similarly, the present work
also yields the underestimated results for the proton and neutron
magnetic moments. As shown in Table~\ref{tab:4}, the valence-quark
contributions are dominant over the sea-quark contributions for both
the proton and neutron magnetic moments. In Table~\ref{tab:5}, the
results for the magnetization radii of the proton and neutron are
presented. Both the results are comparable with the experimental data. 

\begin{figure}[htp]
  \centering
  \includegraphics[scale=0.18]{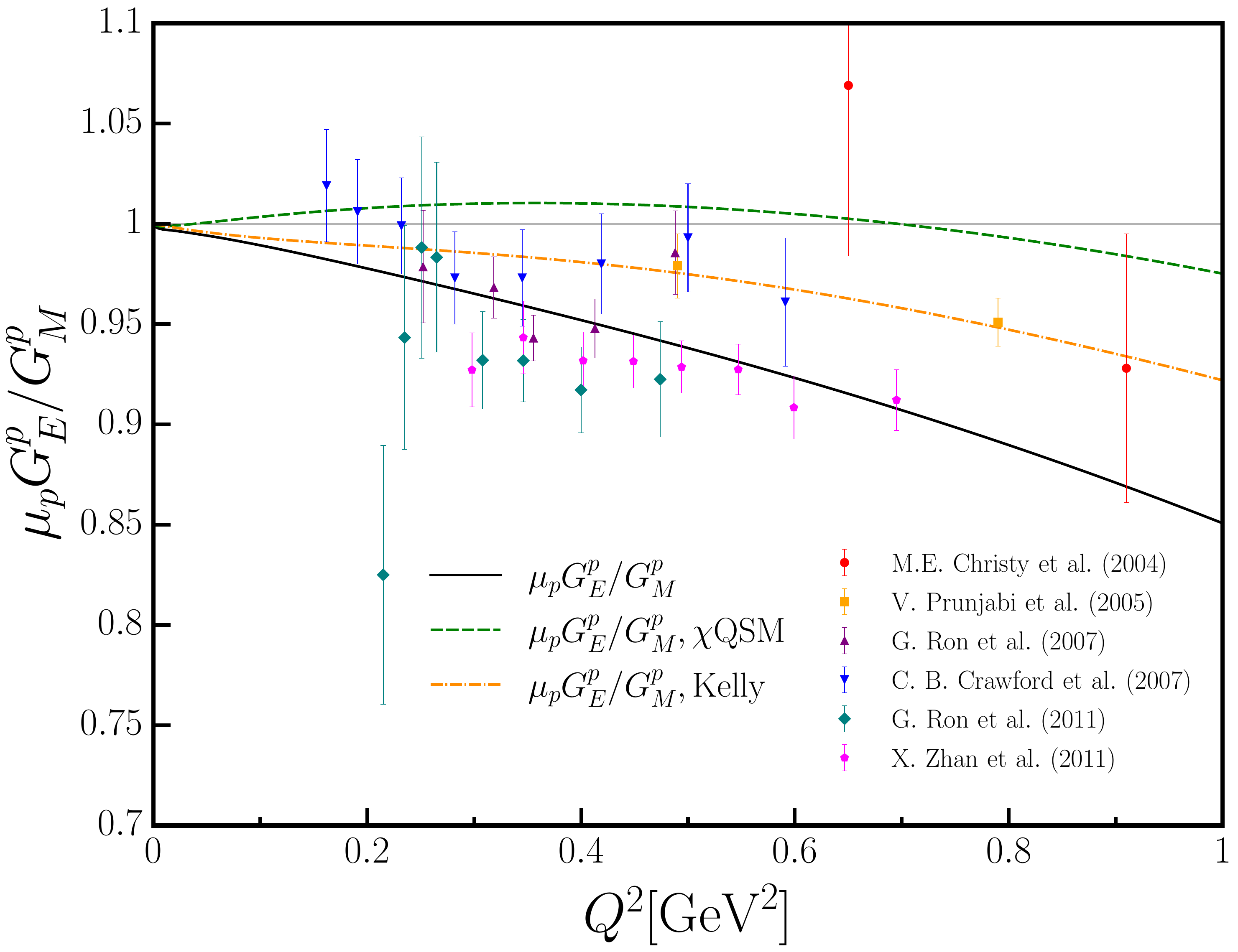}
\caption[Ratio of the electric and magnetic form factors of the proton]
{Ratio of the electric and magnetic form factors $G_E$ and $G_M$ for
  the proton normalized by its magnetic moment. The solid curve
  depicts the present numerical result for the ratio, while the
  dashed and dot-dashed ones represent those obtained from the
  $\chi$QSM with $M_0=420$
  MeV~\cite{Christov:1995hr,Christov:1995vm} and from the Kelly
  parametrization~\cite{Kelly:2004hm}, respectively. The experimental
  data are taken from
  Refs.~\cite{Ron:2007vr,JeffersonLabHallA:2011yyi,Zhan:2011ji,Punjabi:2005wq,E94110:2004lsx,Crawford:2006rz}.}
\label{fig:7}
\end{figure}
\begin{figure}[htp]
  \centering
  \includegraphics[scale=0.18]{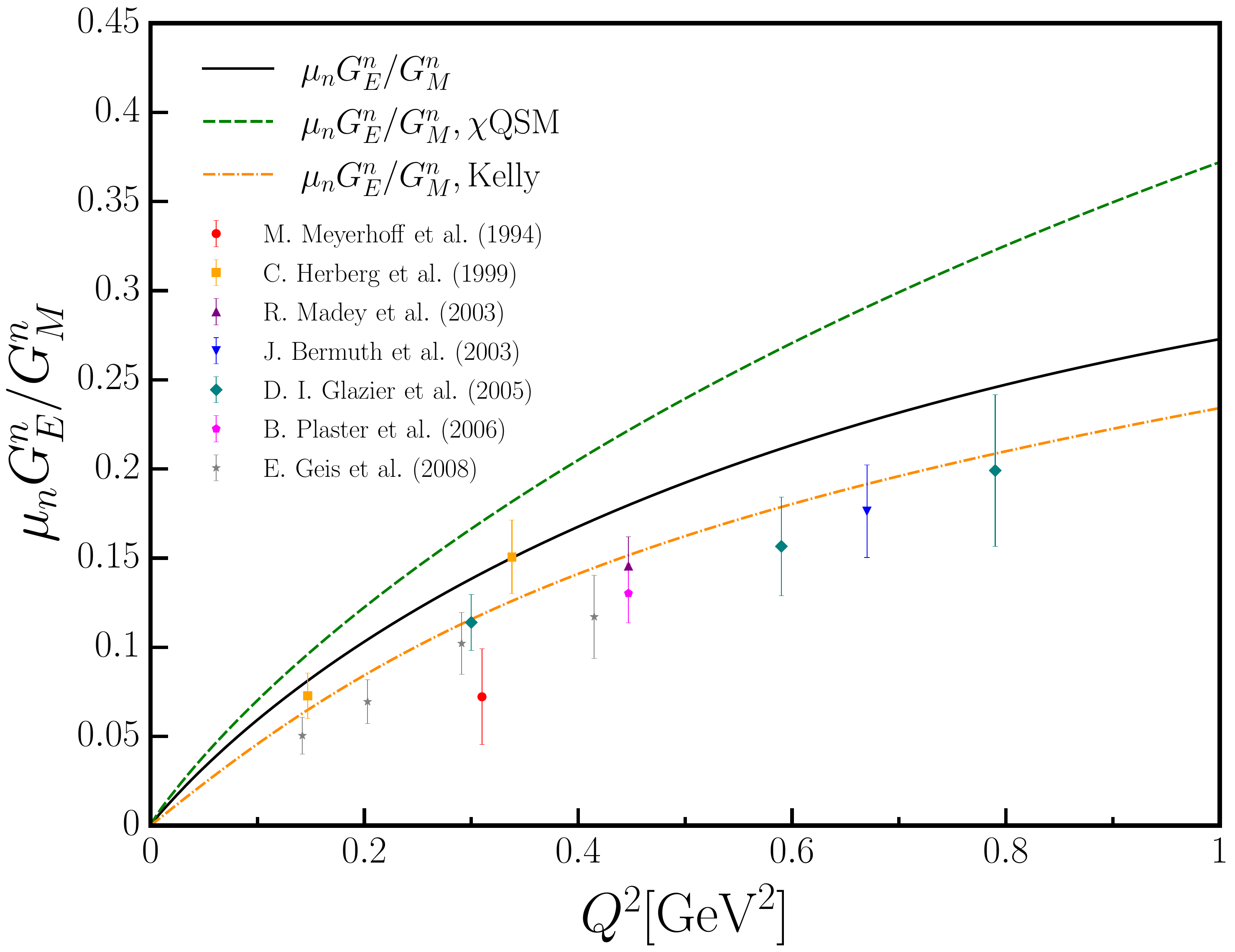}
  \caption[Ratio of the electric and magnetic form factors of the neutron]
  {Ratio of the electric and magnetic form factors $G_E$ and $G_M$ for
    the neutron normalized by its magnetic moment. The solid curve
    depicts the present numerical result for the ratio, while the
    dashed and dot-dashed ones represent those obtained from the
    $\chi$QSM with $M_0=420$
    MeV~\cite{Christov:1995hr,Christov:1995vm} and from the Kelly
    parametrization~\cite{Kelly:2004hm}, respectively. The
    experimental data are taken from
    Refs.~\cite{Meyerhoff:1994ev,Herberg:1999ud,E93-038:2003ixb,Bermuth:2003qh,Glazier:2004ny,JeffersonLaboratoryE93-038:2005ryd,BLAST:2008bub}.}
\label{fig:8}
\end{figure}

The experimental data on the EM form factors of the nucleon used for
comparison with the present results are based on the Rosenbluth
separation method~\cite{Rosenbluth:1950yq}, in which the electric and
magnetic form factors are extracted separately from the differential
cross section of unpolarized electron-proton elastic scattering. With
the advent of high-quality polarized electron beams, however, it
became possible to extract the electric-to-magnetic form-factor ratio
of the proton directly from polarization
observables~\cite{JeffersonLabHallA:1999epl}. This ratio carries
profound physical implications, since it reflects how differently the
electric charge and magnetization are distributed inside the nucleon.
As shown in Ref.~\cite{Ron:2007vr}, $\mu_p G_E^p(Q^2)/G_M^p(Q^2)$
exhibits a clear decrease with increasing $Q^2$. As displayed in
Fig.~\ref{fig:7}, the present result reproduces this behavior very
well, whereas that from the $\chi$QSM first rises slowly and then
falls off only mildly. This indicates that the effective chiral
theory properly accounts for the spatial difference between the
charge and magnetization distributions inside the nucleon. In
Fig.~\ref{fig:8}, the result for the corresponding neutron ratio
appears to be slightly overestimated relative to the data, but still
describes them better than the $\chi$QSM does.

\section{Summary and conclusions}
\label{sec:6}
In the present work, we have investigated the electromagnetic form
factors of the nucleon on the basis of an effective chiral theory
derived from the QCD instanton vacuum, with the finite current quark
mass taken into account. Starting from the improved effective
low-energy QCD partition function beyond the chiral limit, the
instanton parameters were fixed to $\bar{\rho}=0.35$ fm and
$\bar{R}=0.86$ fm, and the dynamical quark mass at zero virtuality
was determined to be $M_0=385$ MeV from the saddle-point equation.
Together with the average current quark mass $\overline{m}=5$ MeV
that yields the physical pion mass, all dynamical parameters were
fixed without any free adjustment. The momentum-dependent dynamical
quark mass arising from the instanton vacuum plays the role of a
regulator, so that no additional regularization scheme is required
to tame the divergences arising from quark loops.

We first computed the classical nucleon mass and the moment of
inertia of the nucleon, and compared them with those obtained within
the chiral quark-soliton model and in the chiral limit. The present
framework yields 
$M_{\mathrm{cl}}=1017$ MeV and $\mathcal{I}=1.164$ fm, from which the
$\Delta$--$N$ mass difference was obtained as $M_{\Delta-N}=254.2$
MeV. We then evaluated the Sachs electric and magnetic form factors of
the proton and neutron, and compared them with the experimental data, the
 chiral quark-soliton model, and the Kelly parametrization. The
 isoscalar electric form factor was found to be larger than the
 isovector one over the entire $Q^2$ range considered, as expected
 from the $N_c$ counting. The proton electric form factor was shown to
 be in remarkable agreement with the experimental data in the lower
 $Q^2$ region ($Q^2 \le 0.4\,\mathrm{GeV}^2$), while it falls off
 faster than the data as $Q^2$ increases. The neutron electric form
 factor was found to be in very good agreement with the experimental
 data and to describe them noticeably better than the chiral
 quark-soliton model. The proton charge radius and the neutron
 mean-square charge radius were 
 obtained as $\sqrt{\langle r^2 \rangle_\mathrm{ch}^p}=0.841$ fm and 
$\langle r^2 \rangle_\mathrm{ch}^n=-0.168\,\mathrm{fm}^2$,
respectively.

For the magnetic form factors, the isovector contribution was found
to dominate over the isoscalar one, as expected from the $N_c$
counting. The proton magnetic form factor normalized by its magnetic
moment describes the experimental data well in the lower $Q^2$
region, while it falls off faster than the data as $Q^2$ increases.
The normalized neutron magnetic form factor turned out to be very
similar to that obtained within the chiral quark-soliton model. The
magnetic moments of the proton and neutron were obtained as
$\mu_p=2.042$ and $\mu_n=-1.307$ in units of the nuclear magneton, in
which the valence-quark contributions dominate over the sea-quark 
contributions. The magnetization radii of the proton and neutron
were also presented and compared with the experimental data.

Finally, we computed the ratios of the electric and magnetic form
factors, $\mu_{p,n} G_E^{p,n}(Q^2)/G_M^{p,n}(Q^2)$, and compared them
with the polarization-transfer data, the chiral quark-soliton model,
and the Kelly parametrization. The proton ratio was shown to reproduce
the decreasing behavior with increasing $Q^2$ observed in the 
experimental data. The neutron ratio was found to be slightly
overestimated relative to the data.

Last but not least, we want to mention that the present effective
chiral theory for the structure of the nucleon possesses a natural
normalization point given by the inverse of the average instanton
size~\cite{Diakonov:1985eg,Kim:1995bq,Diakonov:1995qy,Polyakov:2020cnc},
i.e., $\bar{\rho}^{-1}\approx 0.6\,\mathrm{GeV}$. Interestingly, the
present results describe the experimental data very well in the
region $Q^2 \le \bar{\rho}^{-2}$. This indicates that the present
theoretical framework is well suited for describing the
electromagnetic structure of the nucleon. An important advantage of
the present framework is that gluonic observables can be directly
computed within
it~\cite{Diakonov:1995qy,Kim:2024cbq}. Corresponding studies are
currently under way.

\section*{Acknowledgments}
The present work was supported by the National Research Foundation of
Korea (NRF) grant funded by the Korea government under Grant
RS-2025-00513982 (HChK and HJL) and No. RS-2025-02634319 (YC). 

\bibliography{EMFF}

@article{Hofstadter:1956qs,
    author = "Hofstadter, Robert",
    title = "{Electron scattering and nuclear structure}",
    doi = "10.1103/RevModPhys.28.214",
    journal = "Rev. Mod. Phys.",
    volume = "28",
    pages = "214--254",
    year = "1956"
}

@article{Sachs:1962zzc,
    author = "Sachs, R. G.",
    title = "{High-Energy Behavior of Nucleon Electromagnetic Form Factors}",
    doi = "10.1103/PhysRev.126.2256",
    journal = "Phys. Rev.",
    volume = "126",
    pages = "2256--2260",
    year = "1962"
}

@article{Choi:2025xha,
    author = "Choi, Yongwoo and Kim, Hyun-Chul",
    title = "{Nucleon and singly heavy baryons from the QCD instanton vacuum}",
    eprint = "2501.12114",
    archivePrefix = "arXiv",
    primaryClass = "hep-ph",
    reportNumber = "INHA-NTG-01/2025",
    doi = "10.1103/PhysRevD.111.074023",
    journal = "Phys. Rev. D",
    volume = "111",
    number = "7",
    pages = "074023",
    year = "2025"
}

@article{Musakhanov:2002vu,
    author = "Musakhanov, M.",
    editor = "Boffi, S. and Ciofi degli Atti, Claudio and Giannini, M.",
    title = "{Current mass dependence of the quark condensate in instanton vacuum}",
    doi = "10.1016/S0375-9474(01)01516-0",
    journal = "Nucl. Phys. A",
    volume = "699",
    pages = "340--343",
    year = "2002"
}

@article{Goeke:2007bj,
    author = "Goeke, K. and Musakhanov, M. M. and Siddikov, M.",
    title = "{Low energy constants of chi PT from the instanton vacuum model}",
    eprint = "0707.1997",
    archivePrefix = "arXiv",
    primaryClass = "hep-ph",
    doi = "10.1103/PhysRevD.76.076007",
    journal = "Phys. Rev. D",
    volume = "76",
    pages = "076007",
    year = "2007"
}

@article{Broniowski:2001cx,
    author = "Broniowski, Wojciech and Golli, Bojan and Ripka, Georges",
    title = "{Solitons in nonlocal chiral quark models}",
    eprint = "hep-ph/0107139",
    archivePrefix = "arXiv",
    doi = "10.1016/S0375-9474(01)01670-0",
    journal = "Nucl. Phys. A",
    volume = "703",
    pages = "667--701",
    year = "2002"
}

@article{Kahana:1984be,
    author = "Kahana, S. and Ripka, G.",
    title = "{Baryon Density of Quarks Coupled to a Chiral Field}",
    doi = "10.1016/0375-9474(84)90692-4",
    journal = "Nucl. Phys. A",
    volume = "429",
    pages = "462--476",
    year = "1984"
}

@article{Musakhanov:2002xa,
    author = "Musakhanov, M. M. and Kim, Hyun-Chul",
    title = "{A Test of the instanton vacuum with low-energy theorems of the axial anomaly}",
    eprint = "hep-ph/0206233",
    archivePrefix = "arXiv",
    reportNumber = "PNU-NTG-02-2002",
    doi = "10.1016/j.physletb.2003.08.022",
    journal = "Phys. Lett. B",
    volume = "572",
    pages = "181--188",
    year = "2003"
}

@article{Yennie:1957skg,
    author = "Yennie, D. R. and L{\'e}vy, M. M. and Ravenhall, D. G.",
    title = "{Electromagnetic Structure of Nucleons}",
    doi = "10.1103/RevModPhys.29.144",
    journal = "Rev. Mod. Phys.",
    volume = "29",
    number = "1",
    pages = "144",
    year = "1957"
}

@article{Miller:2007uy,
    author = "Miller, Gerald A.",
    title = "{Charge Density of the Neutron}",
    eprint = "0705.2409",
    archivePrefix = "arXiv",
    primaryClass = "nucl-th",
    reportNumber = "NT@UW-07-07",
    doi = "10.1103/PhysRevLett.99.112001",
    journal = "Phys. Rev. Lett.",
    volume = "99",
    pages = "112001",
    year = "2007"
}

@article{Lorce:2020onh,
    author = "Lorc{\'e}, C{\'e}dric",
    title = "{Charge Distributions of Moving Nucleons}",
    eprint = "2007.05318",
    archivePrefix = "arXiv",
    primaryClass = "hep-ph",
    doi = "10.1103/PhysRevLett.125.232002",
    journal = "Phys. Rev. Lett.",
    volume = "125",
    number = "23",
    pages = "232002",
    year = "2020"
}

@article{ParticleDataGroup:2024cfk,
    author = "Navas, S. and others",
    collaboration = "Particle Data Group",
    title = "{Review of particle physics}",
    doi = "10.1103/PhysRevD.110.030001",
    journal = "Phys. Rev. D",
    volume = "110",
    number = "3",
    pages = "030001",
    year = "2024"
}

@article{ParticleDataGroup:2010dbb,
    author = "Nakamura, K. and others",
    collaboration = "Particle Data Group",
    title = "{Review of particle physics}",
    reportNumber = "FERMILAB-PUB-10-665-PPD",
    doi = "10.1088/0954-3899/37/7A/075021",
    journal = "J. Phys. G",
    volume = "37",
    pages = "075021",
    year = "2010"
}

@article{Bartel:1973rf,
    author = "Bartel, W. and Busser, F. W. and Dix, W. r. and Felst, R. and Harms, D. and Krehbiel, H. and Kuhlmann, P. E. and McElroy, J. and Meyer, J. and Weber, G.",
    title = "{Measurement of proton and neutron electromagnetic form-factors at squared four momentum transfers up to 3-GeV/c$^2$}",
    reportNumber = "DESY-73-05",
    doi = "10.1016/0550-3213(73)90594-4",
    journal = "Nucl. Phys. B",
    volume = "58",
    pages = "429--475",
    year = "1973"
}

@article{E94110:2004lsx,
    author = "Christy, M. E. and others",
    collaboration = "E94110",
    title = "{Measurements of electron proton elastic cross-sections for 0.4 {\ensuremath{<}} Q**2 {\ensuremath{<}} 5.5 (GeV/c)**2}",
    eprint = "nucl-ex/0401030",
    archivePrefix = "arXiv",
    reportNumber = "JLAB-PHY-04-216",
    doi = "10.1103/PhysRevC.70.015206",
    journal = "Phys. Rev. C",
    volume = "70",
    pages = "015206",
    year = "2004"
}

@article{Arrington:2007ux,
    author = "Arrington, J. and Melnitchouk, W. and Tjon, J. A.",
    title = "{Global analysis of proton elastic form factor data with two-photon exchange corrections}",
    eprint = "0707.1861",
    archivePrefix = "arXiv",
    primaryClass = "nucl-ex",
    reportNumber = "JLAB-THY-07-678",
    doi = "10.1103/PhysRevC.76.035205",
    journal = "Phys. Rev. C",
    volume = "76",
    pages = "035205",
    year = "2007"
}

@article{Arrington:2015ria,
    author = "Arrington, John and Sick, Ingo",
    title = "{Evaluation of the proton charge radius from e-p scattering}",
    eprint = "1505.02680",
    archivePrefix = "arXiv",
    primaryClass = "nucl-ex",
    doi = "10.1063/1.4921430",
    journal = "J. Phys. Chem. Ref. Data",
    volume = "44",
    pages = "031204",
    year = "2015"
}

@article{A1:2013fsc,
    author = "Bernauer, J. C. and others",
    collaboration = "A1",
    title = "{Electric and magnetic form factors of the proton}",
    eprint = "1307.6227",
    archivePrefix = "arXiv",
    primaryClass = "nucl-ex",
    doi = "10.1103/PhysRevC.90.015206",
    journal = "Phys. Rev. C",
    volume = "90",
    number = "1",
    pages = "015206",
    year = "2014"
}

@article{Herberg:1999ud,
    author = "Herberg, C. and others",
    title = "{Determination of the neutron electric form-factor in the D(e,e' n)p reaction and the influence of nuclear binding}",
    doi = "10.1007/s100500050268",
    journal = "Eur. Phys. J. A",
    volume = "5",
    pages = "131--135",
    year = "1999"
}

@article{Passchier:1999cj,
    author = "Passchier, I. and others",
    title = "{The Charge form-factor of the neutron from the reaction polarized H-2(polarized e, e-prime n) p}",
    eprint = "nucl-ex/9907012",
    archivePrefix = "arXiv",
    reportNumber = "JLAB-PHY-99-67",
    doi = "10.1103/PhysRevLett.82.4988",
    journal = "Phys. Rev. Lett.",
    volume = "82",
    pages = "4988--4991",
    year = "1999"
}

@article{Schiavilla:2001qe,
    author = "Schiavilla, R. and Sick, I.",
    title = "{Neutron charge form-factor at large q**2}",
    eprint = "nucl-ex/0107004",
    archivePrefix = "arXiv",
    reportNumber = "JLAB-THY-01-23",
    doi = "10.1103/PhysRevC.64.041002",
    journal = "Phys. Rev. C",
    volume = "64",
    pages = "041002",
    year = "2001"
}

@article{JeffersonLabE93-026:2003tty,
    author = "Warren, G. and others",
    collaboration = "Jefferson Lab E93-026",
    title = "{Measurement of the electric form-factor of the neutron at $Q^2$ = 0.5 and 1.0 $GeV^2/c^2$}",
    eprint = "nucl-ex/0308021",
    archivePrefix = "arXiv",
    reportNumber = "JLAB-PHY-03-215",
    doi = "10.1103/PhysRevLett.92.042301",
    journal = "Phys. Rev. Lett.",
    volume = "92",
    pages = "042301",
    year = "2004"
}

@article{JeffersonLaboratoryE93-038:2005ryd,
    author = "Plaster, B. and others",
    collaboration = "Jefferson Laboratory E93-038",
    title = "{Measurements of the neutron electric to magnetic form-factor ratio G(En) / G(Mn) via the H-2(polarized-e, e-prime,polarized-n)H-1 reaction to Q**2 = 1.45-(GeV/c)**2}",
    eprint = "nucl-ex/0511025",
    archivePrefix = "arXiv",
    reportNumber = "JLAB-PHY-06-504",
    doi = "10.1103/PhysRevC.73.025205",
    journal = "Phys. Rev. C",
    volume = "73",
    pages = "025205",
    year = "2006"
}

@article{E93026:2001css,
    author = "Zhu, H. and others",
    collaboration = "E93026",
    title = "{A Measurement of the electric form-factor of the neutron through polarized-d (polarized-e, e-prime n)p at Q**2 = 0.5-(GeV/c)**2}", 
    eprint = "nucl-ex/0105001",
    archivePrefix = "arXiv",
    reportNumber = "JLAB-PHY-01-45",
    doi = "10.1103/PhysRevLett.87.081801",
    journal = "Phys. Rev. Lett.",
    volume = "87",
    pages = "081801",
    year = "2001"
}

@article{Pohl:2010zza,
    author = "Pohl, Randolf and others",
    title = "{The size of the proton}",
    doi = "10.1038/nature09250",
    journal = "Nature",
    volume = "466",
    pages = "213--216",
    year = "2010"
}

@article{Pohl:2013yb,
    author = "Pohl, Randolf and Gilman, Ronald and Miller, Gerald A. and Pachucki, Krzysztof",
    title = "{Muonic hydrogen and the proton radius puzzle}",
    eprint = "1301.0905",
    archivePrefix = "arXiv",
    primaryClass = "physics.atom-ph",
    doi = "10.1146/annurev-nucl-102212-170627",
    journal = "Ann. Rev. Nucl. Part. Sci.",
    volume = "63",
    pages = "175--204",
    year = "2013"
}

@article{Antognini:2013txn,
    author = "Antognini, Aldo and others",
    title = "{Proton Structure from the Measurement of $2S-2P$ Transition Frequencies of Muonic Hydrogen}",
    doi = "10.1126/science.1230016",
    journal = "Science",
    volume = "339",
    pages = "417--420",
    year = "2013"
}

@article{Antognini:2022xoo,
    author = "Antognini, Aldo and Hagelstein, Franziska and Pascalutsa, Vladimir",
    title = "{The proton structure in and out of muonic hydrogen}",
    eprint = "2205.10076",
    archivePrefix = "arXiv",
    primaryClass = "nucl-th",
    reportNumber = "MITP/22-039, PSI-PR-22-29",
    doi = "10.1146/annurev-nucl-101920-024709",
    journal = "Ann. Rev. Nucl. Part. Sci.",
    volume = "72",
    pages = "389",
    year = "2022"
}

@article{Higinbotham:2015rja,
    author = "Higinbotham, Douglas W. and Kabir, Al Amin and Lin, Vincent and Meekins, David and Norum, Blaine and Sawatzky, Brad",
    title = "{Proton radius from electron scattering data}",
    eprint = "1510.01293",
    archivePrefix = "arXiv",
    primaryClass = "nucl-ex",
    reportNumber = "JLAB-PHY-16-2",
    doi = "10.1103/PhysRevC.93.055207",
    journal = "Phys. Rev. C",
    volume = "93",
    number = "5",
    pages = "055207",
    year = "2016"
}

@article{Xiong:2019umf,
    author = "Xiong, W. and others",
    title = "{A small proton charge radius from an electron{\textendash}proton scattering experiment}",
    doi = "10.1038/s41586-019-1721-2",
    journal = "Nature",
    volume = "575",
    number = "7781",
    pages = "147--150",
    year = "2019"
}

@article{Maisenbacher:2026nau,
    author = {Maisenbacher, Lothar and Wirthl, Vitaly and Matveev, Arthur and Grinin, Alexey and Pohl, Randolf and H{\"a}nsch, Theodor W. and Udem, Thomas},
    title = "{Sub-part-per-trillion test of the Standard Model with atomic hydrogen}",
    eprint = "2602.14980",
    archivePrefix = "arXiv",
    primaryClass = "physics.atom-ph",
    doi = "10.1038/s41586-026-10124-3",
    journal = "Nature",
    volume = "650",
    number = "8103",
    pages = "845--851",
    year = "2026"
}

@article{Mohr:2024kco,
    author = "Mohr, Peter J. and Newell, David B. and Taylor, Barry N. and Tiesinga, Eite",
    title = "{CODATA recommended values of the fundamental physical constants: 2022*}",
    eprint = "2409.03787",
    archivePrefix = "arXiv",
    primaryClass = "hep-ph",
    doi = "10.1103/RevModPhys.97.025002",
    journal = "Rev. Mod. Phys.",
    volume = "97",
    number = "2",
    pages = "025002",
    year = "2025"
}

@article{Scheidegger:2024rrm,
    author = "Scheidegger, Simon and Merkt, Fr{\'e}d{\'e}ric",
    title = "{Precision-Spectroscopic Determination of the Binding Energy of a Two-Body Quantum System: The Hydrogen Atom and the Proton-Size Puzzle}",
    doi = "10.1103/PhysRevLett.132.113001",
    journal = "Phys. Rev. Lett.",
    volume = "132",
    number = "11",
    pages = "113001",
    year = "2024"
}

@article{Brandt:2021yor,
    author = "Brandt, A. D. and Cooper, S. F. and Rasor, C. and Burkley, Z. and Yost, D. C. and Matveev, A.",
    title = "{Measurement of the 2S1/2-8D5/2 Transition in Hydrogen}",
    eprint = "2111.08554",
    archivePrefix = "arXiv",
    primaryClass = "physics.atom-ph",
    doi = "10.1103/PhysRevLett.128.023001",
    journal = "Phys. Rev. Lett.",
    volume = "128",
    number = "2",
    pages = "023001",
    year = "2022"
}

@article{Grinin:2020txk,
    author = {Grinin, Alexey and Matveev, Arthur and Yost, Dylan C. and Maisenbacher, Lothar and Wirthl, Vitaly and Pohl, Randolf and H{\"a}nsch, Theodor W. and Udem, Thomas},
    title = "{Two-photon frequency comb spectroscopy of atomic hydrogen}",
    doi = "10.1126/science.abc7776",
    journal = "Science",
    volume = "370",
    number = "6520",
    pages = "abc7776",
    year = "2020"
}

@article{Fleurbaey:2018fih,
    author = "Fleurbaey, H{\'e}l{\`e}ne and Galtier, Sandrine and Thomas, Simon and Bonnaud, Marie and Julien, Lucile and Biraben, Fran{\c{c}}ois and Nez, Fran{\c{c}}ois and Abgrall, Michel and Gu{\'e}na, Jocelyne",
    title = "{New Measurement of the $1S-3S$ Transition Frequency of Hydrogen: Contribution to the Proton Charge Radius Puzzle}",
    eprint = "1801.08816",
    archivePrefix = "arXiv",
    primaryClass = "physics.atom-ph",
    doi = "10.1103/PhysRevLett.120.183001",
    journal = "Phys. Rev. Lett.",
    volume = "120",
    number = "18",
    pages = "183001",
    year = "2018"
}

@article{Horbatsch:2015qda,
    author = "Horbatsch, M. and Hessels, E. A.",
    title = "{Evaluation of the strength of electron-proton scattering data for determining the proton charge radius}",
    eprint = "1509.05644",
    archivePrefix = "arXiv",
    primaryClass = "nucl-ex",
    doi = "10.1103/PhysRevC.93.015204",
    journal = "Phys. Rev. C",
    volume = "93",
    number = "1",
    pages = "015204",
    year = "2016"
}

@article{Bezginov:2019mdi,
    author = "Bezginov, N. and Valdez, T. and Horbatsch, M. and Marsman, A. and Vutha, A. C. and Hessels, E. A.",
    title = "{A measurement of the atomic hydrogen Lamb shift and the proton charge radius}",
    doi = "10.1126/science.aau7807",
    journal = "Science",
    volume = "365",
    number = "6457",
    pages = "1007--1012",
    year = "2019"
}

@article{Christov:1995hr,
    author = "Christov, C. V. and Gorski, A. Z. and Goeke, K. and Pobylitsa, P. V.",
    title = "{Electromagnetic form-factors of the nucleon in the chiral quark soliton model}",
    eprint = "hep-ph/9507256",
    archivePrefix = "arXiv",
    reportNumber = "RUB-TPII-24-95",
    doi = "10.1016/0375-9474(95)00309-O",
    journal = "Nucl. Phys. A",
    volume = "592",
    pages = "513--538",
    year = "1995"
}

@article{Kim:1995mr,
    author = "Kim, Hyun-Chul and Blotz, Andree and Polyakov, Maxim V. and Goeke, Klaus",
    title = "{Electromagnetic form-factors of the SU(3) Octet baryons in the semibosonized SU(3) Nambu-Jona-Lasinio model}",
    eprint = "hep-ph/9504363",
    archivePrefix = "arXiv",
    reportNumber = "RUB-TPII-7-95",
    doi = "10.1103/PhysRevD.53.4013",
    journal = "Phys. Rev. D",
    volume = "53",
    pages = "4013--4029",
    year = "1996"
}

@article{Walker:1993vj,
    author = "Walker, R. C. and others",
    title = "{Measurements of the proton elastic form-factors for 1-GeV/c**2 {\ensuremath{<}}= Q**2 {\ensuremath{<}}= 3-GeV/C**2 at SLAC}",
    reportNumber = "SLAC-PUB-5815, UR-1305, ER-40685-754",
    doi = "10.1103/PhysRevD.49.5671",
    journal = "Phys. Rev. D",
    volume = "49",
    pages = "5671--5689",
    year = "1994"
}

@article{Anklin:1998ae,
    author = "Anklin, H. and others",
    title = "{Precise measurements of the neutron magnetic form-factor}",
    doi = "10.1016/S0370-2693(98)00442-0",
    journal = "Phys. Lett. B",
    volume = "428",
    pages = "248--253",
    year = "1998"
}

@article{Kubon:2001rj,
    author = "Kubon, G. and others",
    title = "{Precise neutron magnetic form-factors}",
    eprint = "nucl-ex/0107016",
    archivePrefix = "arXiv",
    doi = "10.1016/S0370-2693(01)01386-7",
    journal = "Phys. Lett. B",
    volume = "524",
    pages = "26--32",
    year = "2002"
}

@article{JeffersonLabE95-001:2006dax,
    author = "Anderson, B. and others",
    collaboration = "Jefferson Lab E95-001",
    title = "{Extraction of the Neutron Magnetic Form Factor from Quasi-elastic $^{3}\vec{He}(\vec{e},e')$ at Q$^2$ = 0.1 - 0.6 (GeV/c)$^2$}",
    eprint = "nucl-ex/0605006",
    archivePrefix = "arXiv",
    reportNumber = "JLAB-PHY-06-491",
    doi = "10.1103/PhysRevC.75.034003",
    journal = "Phys. Rev. C",
    volume = "75",
    pages = "034003",
    year = "2007"
}

@article{CLAS:2008idi,
    author = "Lachniet, J. and others",
    collaboration = "CLAS",
    title = "{A Precise Measurement of the Neutron Magnetic Form Factor G**n(M)in the Few-GeV**2 Region}",
    eprint = "0811.1716",
    archivePrefix = "arXiv",
    primaryClass = "nucl-ex",
    reportNumber = "JLAB-PHY-08-911",
    doi = "10.1103/PhysRevLett.102.192001",
    journal = "Phys. Rev. Lett.",
    volume = "102",
    pages = "192001",
    year = "2009"
}

@article{JeffersonLabHallA:2023rsh,
    author = "Santiesteban, S. N. and others",
    collaboration = "Jefferson Lab Hall A",
    title = "{Novel Measurement of the Neutron Magnetic Form Factor from A=3 Mirror Nuclei}",
    eprint = "2304.13770",
    archivePrefix = "arXiv",
    primaryClass = "nucl-ex",
    reportNumber = "FERMILAB-PUB-23-195-T",
    doi = "10.1103/PhysRevLett.132.162501",
    journal = "Phys. Rev. Lett.",
    volume = "132",
    number = "16",
    pages = "162501",
    year = "2024"
}

@article{Golak:2000nt,
    author = "Golak, J. and Ziemer, G. and Kamada, H. and Witala, H. and Gloeckle, Walter",
    title = "{Extraction of electromagnetic neutron form-factors through inclusive and exclusive polarized electron scattering on polarized He-3 target}",
    eprint = "nucl-th/0008008",
    archivePrefix = "arXiv",
    doi = "10.1103/PhysRevC.63.034006",
    journal = "Phys. Rev. C",
    volume = "63",
    pages = "034006",
    year = "2001"
}

@article{E93-038:2003ixb,
    author = "Madey, R. and others",
    collaboration = "E93-038",
    title = "{Measurements of G(E)n / G(M)n from the H-2(polarized-e,e-prime polarized-n) reaction to Q**2 = 1.45 (GeV/c)**2}",
    eprint = "nucl-ex/0308007",
    archivePrefix = "arXiv",
    reportNumber = "JLAB-PHY-03-224",
    doi = "10.1103/PhysRevLett.91.122002",
    journal = "Phys. Rev. Lett.",
    volume = "91",
    pages = "122002",
    year = "2003"
}

@article{Perdrisat:2006hj,
    author = "Perdrisat, C. F. and Punjabi, V. and Vanderhaeghen, M.",
    title = "{Nucleon Electromagnetic Form Factors}",
    eprint = "hep-ph/0612014",
    archivePrefix = "arXiv",
    reportNumber = "WM-06-115, JLAB-THY-06-595",
    doi = "10.1016/j.ppnp.2007.05.001",
    journal = "Prog. Part. Nucl. Phys.",
    volume = "59",
    pages = "694--764",
    year = "2007"
}

@article{Arrington:2011kb,
    author = "Arrington, John and de Jager, Kees and Perdrisat, Charles F.",
    title = "{Nucleon Form Factors: A Jefferson Lab Perspective}",
    eprint = "1102.2463",
    archivePrefix = "arXiv",
    primaryClass = "nucl-ex",
    reportNumber = "JLAB-PHY-11-1315",
    doi = "10.1088/1742-6596/299/1/012002",
    journal = "J. Phys. Conf. Ser.",
    volume = "299",
    pages = "012002",
    year = "2011"
}

@article{Ron:2007vr,
    author = "Ron, G. and others",
    title = "{The Proton Elastic Form Factor Ratio mu(p) G**p(E)/G**p(M) at Low Momentum Transfer}",
    eprint = "0706.0128",
    archivePrefix = "arXiv",
    primaryClass = "nucl-ex",
    reportNumber = "JLAB-PHY-07-650",
    doi = "10.1103/PhysRevLett.99.202002",
    journal = "Phys. Rev. Lett.",
    volume = "99",
    pages = "202002",
    year = "2007"
}

@article{JeffersonLabHallA:2011yyi,
    author = "Ron, G. and others",
    collaboration = "Jefferson Lab Hall A",
    title = "{Low $Q^2$ measurements of the proton form factor ratio $mu_p G_E / G_M$}",
    eprint = "1103.5784",
    archivePrefix = "arXiv",
    primaryClass = "nucl-ex",
    reportNumber = "JLAB-PHY-11-1415",
    doi = "10.1103/PhysRevC.84.055204",
    journal = "Phys. Rev. C",
    volume = "84",
    pages = "055204",
    year = "2011"
}

@article{Zhan:2011ji,
    author = "Zhan, X. and others",
    title = "{High-Precision Measurement of the Proton Elastic Form Factor Ratio $\mu_pG_E/G_M$ at low $Q^2$}",
    eprint = "1102.0318",
    archivePrefix = "arXiv",
    primaryClass = "nucl-ex",
    reportNumber = "JLAB-PHY-11-1311",
    doi = "10.1016/j.physletb.2011.10.002",
    journal = "Phys. Lett. B",
    volume = "705",
    pages = "59--64",
    year = "2011"
}

@article{Punjabi:2005wq,
    author = "Punjabi, V. and others",
    title = "{Proton elastic form-factor ratios to Q**2 = 3.5-GeV**2 by polarization transfer}",
    eprint = "nucl-ex/0501018",
    archivePrefix = "arXiv",
    reportNumber = "JLAB-PHY-05-292",
    doi = "10.1103/PhysRevC.71.055202",
    journal = "Phys. Rev. C",
    volume = "71",
    pages = "055202",
    year = "2005",
    note = "[Erratum: Phys.Rev.C 71, 069902 (2005)]"
}

@article{Crawford:2006rz,
    author = "Crawford, Christopher B. and others",
    title = "{Measurement of the proton electric to magnetic form factor ratio from vector H-1(vector e, e' p)}",
    eprint = "nucl-ex/0609007",
    archivePrefix = "arXiv",
    doi = "10.1103/PhysRevLett.98.052301",
    journal = "Phys. Rev. Lett.",
    volume = "98",
    pages = "052301",
    year = "2007"
}

@article{BLAST:2008bub,
    author = "Geis, E. and others",
    collaboration = "BLAST",
    title = "{The Charge Form Factor of the Neutron at Low Momentum Transfer from the H-2-polarized (e-polarized, e-prime n) p Reaction}",
    eprint = "0803.3827",
    archivePrefix = "arXiv",
    primaryClass = "nucl-ex",
    doi = "10.1103/PhysRevLett.101.042501",
    journal = "Phys. Rev. Lett.",
    volume = "101",
    pages = "042501",
    year = "2008"
}

@article{Alexandrou:2018sjm,
    author = "Alexandrou, C. and Bacchio, S. and Constantinou, M. and Finkenrath, J. and Hadjiyiannakou, K. and Jansen, K. and Koutsou, G. and Vaquero Aviles-Casco, A.",
    title = "{Proton and neutron electromagnetic form factors from lattice QCD}",
    eprint = "1812.10311",
    archivePrefix = "arXiv",
    primaryClass = "hep-lat",
    reportNumber = "DESY-18-033, DESY 18-033",
    doi = "10.1103/PhysRevD.100.014509",
    journal = "Phys. Rev. D",
    volume = "100",
    number = "1",
    pages = "014509",
    year = "2019"
}

@article{Alexandrou:2025vto,
    author = "Alexandrou, Constantia and Bacchio, Simone and Koutsou, Giannis and Prasad, Bhavna and Spanoudes, Gregoris",
    title = "{Proton and neutron electromagnetic form factors from lattice QCD in the continuum limit}",
    eprint = "2507.20910",
    archivePrefix = "arXiv",
    primaryClass = "hep-lat",
    month = "7",
    year = "2025"
}

@article{Atac:2021wqj,
    author = "Atac, H. and Constantinou, M. and Meziani, Z. -E and Paolone, M. and Sparveris, N.",
    title = "{Measurement of the neutron charge radius and the role of its constituents}",
    eprint = "2103.10840",
    archivePrefix = "arXiv",
    primaryClass = "nucl-ex",
    doi = "10.1038/s41467-021-22028-z",
    journal = "Nature Commun.",
    volume = "12",
    number = "1",
    pages = "1759",
    year = "2021"
}

@article{Becker:1999tw,
    author = "Becker, J. and others",
    title = "{Determination of the neutron electric form-factor from the reaction He-3(e,e' n) at medium momentum transfer}",
    doi = "10.1007/s100500050351",
    journal = "Eur. Phys. J. A",
    volume = "6",
    pages = "329--344",
    year = "1999"
}

@article{Bermuth:2003qh,
    author = "Bermuth, J. and others",
    title = "{The Neutron charge form-factor and target analyzing powers from polarized-He-3 (polarized-e,e-prime n) scattering}",
    eprint = "nucl-ex/0303015",
    archivePrefix = "arXiv",
    doi = "10.1016/S0370-2693(03)00725-1",
    journal = "Phys. Lett. B",
    volume = "564",
    pages = "199--204",
    year = "2003"
}

@article{Rohe:1999sh,
    author = "Rohe, D. and others",
    title = "{Measurement of the neutron electric form-factor G(en) at 0.67-(GeV/c)**2 via He-3(pol.)(e(pol.),e' n)}",
    doi = "10.1103/PhysRevLett.83.4257",
    journal = "Phys. Rev. Lett.",
    volume = "83",
    pages = "4257--4260",
    year = "1999"
}

@article{Ostrick:1999xa,
    author = "Ostrick, M. and others",
    title = "{Measurement of the neutron electric form-factor G(E,n) in the quasifree H-2(e(pol.),e' n(pol.))p reaction}",
    doi = "10.1103/PhysRevLett.83.276",
    journal = "Phys. Rev. Lett.",
    volume = "83",
    pages = "276--279",
    year = "1999"
}

@article{Kelly:2004hm,
    author = "Kelly, J. J.",
    title = "{Simple parametrization of nucleon form factors}",
    doi = "10.1103/PhysRevC.70.068202",
    journal = "Phys. Rev. C",
    volume = "70",
    pages = "068202",
    year = "2004"
}

@article{Jang:2019jkn,
    author = "Jang, Yong-Chull and Gupta, Rajan and Lin, Huey-Wen and Yoon, Boram and Bhattacharya, Tanmoy",
    title = "{Nucleon electromagnetic form factors in the continuum limit from ( 2+1+1 )-flavor lattice QCD}",
    eprint = "1906.07217",
    archivePrefix = "arXiv",
    primaryClass = "hep-lat",
    reportNumber = "LA-UR-19-25275, MSUHEP-19-006",
    doi = "10.1103/PhysRevD.101.014507",
    journal = "Phys. Rev. D",
    volume = "101",
    number = "1",
    pages = "014507",
    year = "2020"
}

@article{Capitani:2015sba,
    author = {Capitani, S. and Della Morte, M. and Djukanovic, D. and von Hippel, G. and Hua, J. and J{\"a}ger, B. and Knippschild, B. and Meyer, H. B. and Rae, T. D. and Wittig, H.},
    title = "{Nucleon electromagnetic form factors in two-flavor QCD}",
    eprint = "1504.04628",
    archivePrefix = "arXiv",
    primaryClass = "hep-lat",
    reportNumber = "MITP-15-026, HIM-2015-01, CP3-ORIGINS-2015-012, DIAS-2015-12",
    doi = "10.1103/PhysRevD.92.054511",
    journal = "Phys. Rev. D",
    volume = "92",
    number = "5",
    pages = "054511",
    year = "2015"
}

@article{Meyerhoff:1994ev,
    author = "Meyerhoff, M. and others",
    title = "{First measurement of the electric form-factor of the neutron in the exclusive quasielastic scattering of polarized electrons from polarized He-3}",
    doi = "10.1016/0370-2693(94)90718-8",
    journal = "Phys. Lett. B",
    volume = "327",
    pages = "201--207",
    year = "1994"
}

@article{Glazier:2004ny,
    author = "Glazier, D. I. and others",
    title = "{Measurement of the electric form-factor of the neutron at Q**2 = 0.3-(GeV/c)**2 to 0.8-(GeV/c)**2}",
    eprint = "nucl-ex/0410026",
    archivePrefix = "arXiv",
    doi = "10.1140/epja/i2004-10115-8",
    journal = "Eur. Phys. J. A",
    volume = "24",
    pages = "101--109",
    year = "2005"
}

@article{Punjabi:2015bba,
    author = "Punjabi, V. and Perdrisat, C. F. and Jones, M. K. and Brash, E. J. and Carlson, C. E.",
    title = "{The Structure of the Nucleon: Elastic Electromagnetic Form Factors}",
    eprint = "1503.01452",
    archivePrefix = "arXiv",
    primaryClass = "nucl-ex",
    reportNumber = "JLAB-PHY-15-2019",
    doi = "10.1140/epja/i2015-15079-x",
    journal = "Eur. Phys. J. A",
    volume = "51",
    pages = "79",
    year = "2015"
}

@article{Eden:1994ji,
    author = "Eden, T. and others",
    title = "{Electric form factor of the neutron from the $^{2}H(\overrightarrow{e},e’\overrightarrow{n})^{1}H$ reaction at $Q^{2} =$ 0.255 (GeV/c)$^2$}",
    reportNumber = "CEBAF-PR-94-22",
    doi = "10.1103/PhysRevC.50.R1749",
    journal = "Phys. Rev. C",
    volume = "50",
    number = "4",
    pages = "R1749--R1753",
    year = "1994"
}

@article{Alberico:2008sz,
    author = "Alberico, W. M. and Bilenky, S. M. and Giunti, C. and Graczyk, K. M.",
    title = "{Electromagnetic form factors of the nucleon: New Fit and analysis of uncertainties}",
    eprint = "0812.3539",
    archivePrefix = "arXiv",
    primaryClass = "hep-ph",
    doi = "10.1103/PhysRevC.79.065204",
    journal = "Phys. Rev. C",
    volume = "79",
    pages = "065204",
    year = "2009"
}

@article{Qattan:2012zf,
    author = "Qattan, I. A. and Arrington, J.",
    title = "{Flavor decomposition of the nucleon electromagnetic form factors}",
    eprint = "1209.0683",
    archivePrefix = "arXiv",
    primaryClass = "nucl-ex",
    doi = "10.1103/PhysRevC.86.065210",
    journal = "Phys. Rev. C",
    volume = "86",
    pages = "065210",
    year = "2012"
}

@article{Kim:2024cbq,
    author = "Kim, June-Young and Won, Ho-Yeon and Kim, Hyun-Chul and Weiss, Christian",
    title = "{Spin-orbit correlations in the nucleon in the large-Nc limit}",
    eprint = "2403.07186",
    archivePrefix = "arXiv",
    primaryClass = "hep-ph",
    reportNumber = "JLAB-THY-24-4002, INHA-NTG-01/2024",
    doi = "10.1103/PhysRevD.110.054026",
    journal = "Phys. Rev. D",
    volume = "110",
    number = "5",
    pages = "054026",
    year = "2024"
}

@article{Lorce:2014mxa,
    author = "Lorc{\'e}, C.",
    title = "{Spin{\textendash}orbit correlations in the nucleon}",
    eprint = "1401.7784",
    archivePrefix = "arXiv",
    primaryClass = "hep-ph",
    doi = "10.1016/j.physletb.2014.06.068",
    journal = "Phys. Lett. B",
    volume = "735",
    pages = "344--348",
    year = "2014"
}

@article{Gao:2021sml,
    author = "Gao, Haiyan and Vanderhaeghen, Marc",
    title = "{The proton charge radius}",
    eprint = "2105.00571",
    archivePrefix = "arXiv",
    primaryClass = "hep-ph",
    doi = "10.1103/RevModPhys.94.015002",
    journal = "Rev. Mod. Phys.",
    volume = "94",
    number = "1",
    pages = "015002",
    year = "2022"
}

@article{Muller:1994ses,
    author = {M{\"u}ller, Dieter and Robaschik, D. and Geyer, B. and Dittes, F. -M. and Ho{\v{r}}ej{\v{s}}i, J.},
    title = "{Wave functions, evolution equations and evolution kernels from light ray operators of QCD}",
    eprint = "hep-ph/9812448",
    archivePrefix = "arXiv",
    reportNumber = "NTZ-6-91, NTZ-91-6",
    doi = "10.1002/prop.2190420202",
    journal = "Fortsch. Phys.",
    volume = "42",
    pages = "101--141",
    year = "1994"
}

@article{Ji:1996nm,
    author = "Ji, Xiang-Dong",
    title = "{Deeply virtual Compton scattering}",
    eprint = "hep-ph/9609381",
    archivePrefix = "arXiv",
    reportNumber = "UMD-PP-97-26, MIT-CTP-2568",
    doi = "10.1103/PhysRevD.55.7114",
    journal = "Phys. Rev. D",
    volume = "55",
    pages = "7114--7125",
    year = "1997"
}

@article{Radyushkin:1996ru,
    author = "Radyushkin, A. V.",
    title = "{Asymmetric gluon distributions and hard diffractive electroproduction}",
    eprint = "hep-ph/9605431",
    archivePrefix = "arXiv",
    reportNumber = "CEBAF-TH-96-06",
    doi = "10.1016/0370-2693(96)00844-1",
    journal = "Phys. Lett. B",
    volume = "385",
    pages = "333--342",
    year = "1996"
}

@article{Diehl:2003ny,
    author = "Diehl, M.",
    title = "{Generalized parton distributions}",
    eprint = "hep-ph/0307382",
    archivePrefix = "arXiv",
    reportNumber = "DESY-THESIS-2003-018",
    doi = "10.1016/j.physrep.2003.08.002",
    journal = "Phys. Rept.",
    volume = "388",
    pages = "41--277",
    year = "2003"
}

@article{Goeke:2001tz,
    author = "Goeke, K. and Polyakov, Maxim V. and Vanderhaeghen, M.",
    title = "{Hard exclusive reactions and the structure of hadrons}",
    eprint = "hep-ph/0106012",
    archivePrefix = "arXiv",
    doi = "10.1016/S0146-6410(01)00158-2",
    journal = "Prog. Part. Nucl. Phys.",
    volume = "47",
    pages = "401--515",
    year = "2001"
}

@article{Kim:2021kum,
    author = "Kim, June-Young and Kim, Hyun-Chul",
    title = "{Transverse charge distributions of the nucleon and their Abel images}",
    eprint = "2106.10986",
    archivePrefix = "arXiv",
    primaryClass = "hep-ph",
    reportNumber = "INHA-NTG-05/2021",
    doi = "10.1103/PhysRevD.104.074003",
    journal = "Phys. Rev. D",
    volume = "104",
    number = "7",
    pages = "074003",
    year = "2021"
}

@article{Leader:2013jra,
    author = "Leader, E. and Lorc{\'e}, C.",
    title = "{The angular momentum controversy: What{\textquoteright}s it all about and does it matter?}",
    eprint = "1309.4235",
    archivePrefix = "arXiv",
    primaryClass = "hep-ph",
    doi = "10.1016/j.physrep.2014.02.010",
    journal = "Phys. Rept.",
    volume = "541",
    number = "3",
    pages = "163--248",
    year = "2014"
}

@article{Liu:2025ldh,
  author        = {Liu, Wei-Yang},
  title         = {{Generic framework for non-perturbative QCD in light hadrons}},
  eprint        = {2501.07776},
  archiveprefix = {arXiv},
  primaryclass  = {hep-ph},
  journal       = {},
  month         = {1},
  year          = {2025}
}

@article{Adkins:1983ya,
  author       = {Adkins, Gregory S. and Nappi, Chiara R. and Witten, Edward},
  title        = {{Static Properties of Nucleons in the Skyrme Model}},
  reportnumber = {PRINT-83-0493 (IAS,PRINCETON)},
  doi          = {10.1016/0550-3213(83)90559-X},
  journal      = {Nucl. Phys. B},
  volume       = {228},
  pages        = {552},
  year         = {1983}
}

@article{Diakonov:1985eg,
  author       = {Diakonov, Dmitri and Petrov, V. Yu.},
  title        = {{A Theory of Light Quarks in the Instanton Vacuum}},
  reportnumber = {LENINGRAD-85-1053},
  doi          = {10.1016/0550-3213(86)90011-8},
  journal      = {Nucl. Phys. B},
  volume       = {272},
  pages        = {457--489},
  year         = {1986}
}

@article{Chretien:1954we,
  author  = {Chretien, M. and Peierls, R. E.},
  title   = {{A study of gauge-invariant non-local interactions}},
  doi     = {10.1098/rspa.1954.0131},
  journal = {Proc. Roy. Soc. Lond. A},
  volume  = {223},
  number  = {1155},
  pages   = {468--481},
  year    = {1954}
}

@article{Golli:1998rf,
  author        = {Golli, Bojan and Broniowski, Wojciech and Ripka, Georges},
  title         = {{Solitons in a chiral quark model with nonlocal interactions}},
  eprint        = {hep-ph/9807261},
  archiveprefix = {arXiv},
  reportnumber  = {SACLAY-SPH-T-98-146, INP-1801-PH},
  doi           = {10.1016/S0370-2693(98)00942-3},
  journal       = {Phys. Lett. B},
  volume        = {437},
  pages         = {24--28},
  year          = {1998}
}

@article{Shifman:1978bx,
  author       = {Shifman, Mikhail A. and Vainshtein, A. I. and Zakharov, Valentin I.},
  title        = {{QCD and Resonance Physics. Theoretical Foundations}},
  reportnumber = {ITEP-73-1978, ITEP-80-1978},
  doi          = {10.1016/0550-3213(79)90022-1},
  journal      = {Nucl. Phys. B},
  volume       = {147},
  pages        = {385--447},
  year         = {1979}
}

@article{Witten:1979kh,
  author       = {Witten, Edward},
  title        = {{Baryons in the 1/n Expansion}},
  reportnumber = {HUTP-79-A007},
  doi          = {10.1016/0550-3213(79)90232-3},
  journal      = {Nucl. Phys. B},
  volume       = {160},
  pages        = {57--115},
  year         = {1979}
}

@article{Shuryak:1981ff,
  author       = {Shuryak, Edward V.},
  title        = {{The Role of Instantons in Quantum Chromodynamics. 1. Physical Vacuum}},
  reportnumber = {IYF 81-118},
  doi          = {10.1016/0550-3213(82)90478-3},
  journal      = {Nucl. Phys. B},
  volume       = {203},
  pages        = {93},
  year         = {1982}
}

@article{Schafer:1996wv,
  author        = {Sch\"afer, Thomas and Shuryak, Edward V.},
  title         = {{Instantons in QCD}},
  eprint        = {hep-ph/9610451},
  archiveprefix = {arXiv},
  reportnumber  = {DOE-ER-40561-293, INT-96-00-150},
  doi           = {10.1103/RevModPhys.70.323},
  journal       = {Rev. Mod. Phys.},
  volume        = {70},
  pages         = {323--426},
  year          = {1998}
}

@article{Diakonov:2002fq,
  archiveprefix = {arXiv},
  author        = {Diakonov, Dmitri},
  doi           = {10.1016/S0146-6410(03)90014-7},
  eprint        = {hep-ph/0212026},
  journal       = {Prog. Part. Nucl. Phys.},
  pages         = {173--222},
  reportnumber  = {NORDITA-2002-74-HE},
  title         = {{Instantons at work}},
  volume        = {51},
  year          = {2003}
}

@article{Diakonov:1983hh,
  author       = {Diakonov, Dmitri and Petrov, V. Yu.},
  title        = {{Instanton Based Vacuum from Feynman Variational Principle}},
  reportnumber = {LENINGRAD-83-900},
  doi          = {10.1016/0550-3213(84)90432-2},
  journal      = {Nucl. Phys. B},
  volume       = {245},
  pages        = {259--292},
  year         = {1984}
}

@article{Diakonov:1987ty,
  author       = {Diakonov, Dmitri and Petrov, V. Yu. and Pobylitsa, P. V.},
  title        = {{A Chiral Theory of Nucleons}},
  reportnumber = {LENINGRAD-87-1297},
  doi          = {10.1016/0550-3213(88)90443-9},
  journal      = {Nucl. Phys. B},
  volume       = {306},
  pages        = {809},
  year         = {1988}
}

@inproceedings{Diakonov:1997sj,
  author        = {Diakonov, Dmitri},
  title         = {{Chiral quark - soliton model}},
  booktitle     = {{Advanced Summer School on Nonperturbative Quantum Field Physics}},
  eprint        = {hep-ph/9802298},
  archiveprefix = {arXiv},
  reportnumber  = {NORDITA-98-14},
  pages         = {1--55},
  month         = {6},
  year          = {1997}
}

@article{Blotz:1992pw,
  author       = {Blotz, A. and Diakonov, Dmitri and Goeke, K. and Park, N. W. and Petrov, V. and Pobylitsa, P. V.},
  title        = {{The SU(3) Nambu-Jona-Lasinio soliton in the collective quantization formulation}},
  reportnumber = {RUB-TPII-27-92},
  doi          = {10.1016/0375-9474(93)90505-R},
  journal      = {Nucl. Phys. A},
  volume       = {555},
  pages        = {765--792},
  year         = {1993}
}

@article{Christov:1995vm,
  author        = {Christov, Chr. V. and Blotz, A. and Kim, Hyun-Chul and Pobylitsa, P. and Watabe, T. and Meissner, T. and Ruiz Arriola, E. and Goeke, K.},
  title         = {{Baryons as nontopological chiral solitons}},
  eprint        = {hep-ph/9604441},
  archiveprefix = {arXiv},
  reportnumber  = {RUB-TPII-32-95},
  doi           = {10.1016/0146-6410(96)00057-9},
  journal       = {Prog. Part. Nucl. Phys.},
  volume        = {37},
  pages         = {91--191},
  year          = {1996}
}

@article{Witten:1983tw,
  author       = {Witten, Edward},
  title        = {{Global Aspects of Current Algebra}},
  reportnumber = {PRINT-83-0262 (PRINCETON)},
  doi          = {10.1016/0550-3213(83)90063-9},
  journal      = {Nucl. Phys. B},
  volume       = {223},
  pages        = {422--432},
  year         = {1983}
}

@article{Witten:1983tx,
  author       = {Witten, Edward},
  title        = {{Current Algebra, Baryons, and Quark Confinement}},
  reportnumber = {Print-83-0261 (PRINCETON)},
  doi          = {10.1016/0550-3213(83)90064-0},
  journal      = {Nucl. Phys. B},
  volume       = {223},
  pages        = {433--444},
  year         = {1983}
}

@inproceedings{Ball:1993ak,
  author        = {Ball, Richard D. and Ripka, Georges},
  title         = {{The Regularization of the fermion determinant in chiral quark models}},
  booktitle     = {{Conference on Many-Body Physics}},
  eprint        = {hep-ph/9312260},
  archiveprefix = {arXiv},
  reportnumber  = {SACLAY-SPH-T-93-138, CERN-TH-7122-93},
  pages         = {129--140},
  month         = {12},
  year          = {1993}
}

@article{Diakonov:1995qy,
  author        = {Diakonov, Dmitri and Polyakov, Maxim V. and Weiss, C.},
  title         = {{Hadronic matrix elements of gluon operators in the instanton vacuum}},
  eprint        = {hep-ph/9510232},
  archiveprefix = {arXiv},
  reportnumber  = {RUB-TPII-27-95, PNPI-TH-2076},
  doi           = {10.1016/0550-3213(95)00675-3},
  journal       = {Nucl. Phys. B},
  volume        = {461},
  pages         = {539--580},
  year          = {1996}
}

@article{Kim:2005jc,
  author        = {Kim, Hyun-Chul and Musakhanov, M. M. and Siddikov, M.},
  title         = {{Meson-loop contributions to the quark condensate from the instanton vacuum}},
  eprint        = {hep-ph/0508211},
  archiveprefix = {arXiv},
  reportnumber  = {PNU-NTG-05-2005},
  doi           = {10.1016/j.physletb.2005.11.054},
  journal       = {Phys. Lett. B},
  volume        = {633},
  pages         = {701--709},
  year          = {2006}
}

@article{Nam:2007gf,
  author        = {Nam, Seung-il and Kim, Hyun-Chul},
  title         = {{Electromagnetic form factors of the pion and kaon from the instanton vacuum}},
  eprint        = {0709.1745},
  archiveprefix = {arXiv},
  primaryclass  = {hep-ph},
  reportnumber  = {YITP-07-52, PNU-NTG-10-2007, PNU-NURI-10-2007},
  doi           = {10.1103/PhysRevD.77.094014},
  journal       = {Phys. Rev. D},
  volume        = {77},
  pages         = {094014},
  year          = {2008}
}

@article{Kim:2004hd,
  author        = {Kim, Hyun-Chul and Musakhanov, M. and Siddikov, M.},
  title         = {{Magnetic susceptibility of the QCD vacuum}},
  eprint        = {hep-ph/0411181},
  archiveprefix = {arXiv},
  reportnumber  = {PNU-NTG-06-2004},
  doi           = {10.1016/j.physletb.2004.12.080},
  journal       = {Phys. Lett. B},
  volume        = {608},
  pages         = {95--106},
  year          = {2005}
}

@article{Shim:2017wcq,
  author        = {Shim, Sang-In and Kim, Hyun-Chul},
  title         = {{Pion radiative weak decay from the instanton vacuum}},
  eprint        = {1704.03263},
  archiveprefix = {arXiv},
  primaryclass  = {hep-ph},
  reportnumber  = {INHA-NTG-03-2017},
  doi           = {10.1016/j.physletb.2017.07.037},
  journal       = {Phys. Lett. B},
  volume        = {772},
  pages         = {687--693},
  year          = {2017}
}

@article{Shim:2018rwv,
  author        = {Shim, Sang-In and Hosaka, Atsushi and Kim, Hyun-Chul},
  title         = {{Vector and Axial-vector form factors in radiative kaon decay and flavor SU(3) symmetry breaking}},
  eprint        = {1810.06815},
  archiveprefix = {arXiv},
  primaryclass  = {hep-ph},
  reportnumber  = {INHA-NTG-10/2018},
  doi           = {10.1016/j.physletb.2019.06.046},
  journal       = {Phys. Lett. B},
  volume        = {795},
  pages         = {438--445},
  year          = {2019}
}

@article{Bowler:1994ir,
  author        = {Bowler, R. D. and Birse, M. C.},
  title         = {{A Nonlocal, covariant generalization of the NJL model}},
  eprint        = {hep-ph/9407336},
  archiveprefix = {arXiv},
  reportnumber  = {M-C-TH-94-17},
  doi           = {10.1016/0375-9474(94)00481-2},
  journal       = {Nucl. Phys. A},
  volume        = {582},
  pages         = {655--664},
  year          = {1995}
}

@article{Holdom:1990iq,
  author       = {Holdom, B. and Terning, J. and Verbeek, K.},
  title        = {{Chiral lagrangian from quarks with dynamical mass}},
  reportnumber = {NSF-ITP-90-81},
  doi          = {10.1016/0370-2693(90)90700-G},
  journal      = {Phys. Lett. B},
  volume       = {245},
  pages        = {612--618},
  year         = {1990}
}

@article{Holdom:1989jb,
  author       = {Holdom, B. and Terning, J. and Verbeek, K.},
  title        = {{A Nonlocal Model of Chiral Dynamics}},
  reportnumber = {UTPT-89-23},
  doi          = {10.1016/0370-2693(89)90755-7},
  journal      = {Phys. Lett. B},
  volume       = {232},
  pages        = {351--356},
  year         = {1989}
}

@article{Holdom:1992fn,
  author       = {Holdom, B.},
  title        = {{Approaching low-energy QCD with a gauged, nonlocal, constituent quark model}},
  reportnumber = {UTPT-91-20, DPNU-91-34-REV},
  doi          = {10.1103/PhysRevD.45.2534},
  journal      = {Phys. Rev. D},
  volume       = {45},
  pages        = {2534--2541},
  year         = {1992}
}

@article{Bos:1991sz,
  author  = {Bos, J. W. and Koch, J. H. and Naus, H. W. L.},
  title   = {{Currents and Ward-Takahashi identities for nonlocal quantum field theories}},
  doi     = {10.1103/PhysRevC.44.485},
  journal = {Phys. Rev. C},
  volume  = {44},
  number  = {1},
  pages   = {485--490},
  year    = {1991}
}

@inproceedings{Broniowski:1999bz,
  author        = {Broniowski, Wojciech},
  title         = {{Gauging nonlocal quark models}},
  booktitle     = {{Mini-Workshop Bled 1999}: {Hadrons as Solitons}},
  eprint        = {hep-ph/9909438},
  archiveprefix = {arXiv},
  reportnumber  = {INP-1828-PH},
  pages         = {17--26},
  month         = {9},
  year          = {1999}
}

@article{Pobylitsa:1989uq,
  author  = {Pobylitsa, P. V.},
  title   = {{The Quark Propagator and Correlation Functions in the Instanton Vacuum}},
  doi     = {10.1016/0370-2693(89)91216-1},
  journal = {Phys. Lett. B},
  volume  = {226},
  pages   = {387--392},
  year    = {1989}
}

@article{Goeke:2007nc,
    author = "Goeke, Klaus and Kim, Hyun-Chul and Musakhanov, M. M. and Siddikov, Marat",
    title = "{1/N(c) corrections to the magnetic susceptibility of the QCD vacuum}",
    eprint = "0708.3526",
    archivePrefix = "arXiv",
    primaryClass = "hep-ph",
    reportNumber = "PNU-NTG-09-2007",
    doi = "10.1103/PhysRevD.76.116007",
    journal = "Phys. Rev. D",
    volume = "76",
    pages = "116007",
    year = "2007"
}

@article{Wakamatsu:1990ud,
    author = "Wakamatsu, M. and Yoshiki, H.",
    title = "{A Chiral quark model of the nucleon}",
    reportNumber = "PRINT-90-0541 (OSAKA)",
    doi = "10.1016/0375-9474(91)90263-6",
    journal = "Nucl. Phys. A",
    volume = "524",
    pages = "561--600",
    year = "1991"
}

@article{Musakhanov:1998wp,
    author = "Musakhanov, M.",
    title = "{Improved effective action for light quarks beyond chiral limit}",
    eprint = "hep-ph/9810295",
    archivePrefix = "arXiv",
    doi = "10.1007/s100529900017",
    journal = "Eur. Phys. J. C",
    volume = "9",
    pages = "235--243",
    year = "1999"
}

@article{Adkins:1983hy,
    author = "Adkins, Gregory S. and Nappi, Chiara R.",
    title = "{The Skyrme Model with Pion Masses}",
    reportNumber = "Print-83-0704 (IAS,PRINCETON)",
    doi = "10.1016/0550-3213(84)90172-X",
    journal = "Nucl. Phys. B",
    volume = "233",
    pages = "109--115",
    year = "1984"
}

@article{Son:2015bwa,
    author = "Son, Hyeon-Dong and Nam, Seung-il and Kim, Hyun-Chul",
    title = "{Weak $K\to \pi$ generalized form factors and transverse transition quark-spin density from the instanton vacuum}",
    eprint = "1502.01558",
    archivePrefix = "arXiv",
    primaryClass = "hep-ph",
    reportNumber = "INHA-NTG-01-2015, PKNU-NUHATH-2015-01",
    doi = "10.1016/j.physletb.2015.06.036",
    journal = "Phys. Lett. B",
    volume = "747",
    pages = "460--467",
    year = "2015"
}

@article{Nam:2007fx,
    author = "Nam, Seung-il and Kim, Hyun-Chul",
    title = "{Kaon semileptonic decay (K(l3) form factors from the instanton vacuum}",
    eprint = "hep-ph/0703089",
    archivePrefix = "arXiv",
    reportNumber = "PNU-NTG-02-2007, PNU-NURI-02-2007",
    doi = "10.1103/PhysRevD.75.094011",
    journal = "Phys. Rev. D",
    volume = "75",
    pages = "094011",
    year = "2007"
}

@article{Nam:2006sx,
    author = "Nam, Seung-il and Kim, Hyun-Chul",
    title = "{Leading-twist pion and kaon distribution amplitudes in the gauge-invariant nonlocal chiral quark model from the instanton vacuum}",
    eprint = "hep-ph/0609267",
    archivePrefix = "arXiv",
    reportNumber = "PNU-NTG-05-2006, PNU-NURI-06-2006",
    doi = "10.1103/PhysRevD.74.076005",
    journal = "Phys. Rev. D",
    volume = "74",
    pages = "076005",
    year = "2006"
}

@article{Rosenbluth:1950yq,
    author = "Rosenbluth, M. N.",
    title = "{High Energy Elastic Scattering of Electrons on Protons}",
    doi = "10.1103/PhysRev.79.615",
    journal = "Phys. Rev.",
    volume = "79",
    pages = "615--619",
    year = "1950"
}

@article{JeffersonLabHallA:1999epl,
    author = "Jones, M. K. and others",
    collaboration = "Jefferson Lab Hall A",
    title = "{$G_{Ep}/G_{Mp}$ ratio by
polarization transfer in $\vec ep\rightarrow e\vec p$}",
    eprint = "nucl-ex/9910005",
    archivePrefix = "arXiv",
    reportNumber = "JLAB-PHY-00-71",
    doi = "10.1103/PhysRevLett.84.1398",
    journal = "Phys. Rev. Lett.",
    volume = "84",
    pages = "1398--1402",
    year = "2000"
}

@article{Kim:1995bq,
    author = "Kim, Hyun-Chul and Polyakov, Maxim V. and Goeke, Klaus",
    title = "{Nucleon tensor charges in the SU(2) chiral quark - soliton model}",
    eprint = "hep-ph/9509283",
    archivePrefix = "arXiv",
    reportNumber = "RUB-TPII-26-95",
    doi = "10.1103/PhysRevD.53.R4715",
    journal = "Phys. Rev. D",
    volume = "53",
    pages = "4715--4718",
    year = "1996"
}

@article{Polyakov:2020cnc,
    author = "Polyakov, Maxim V. and Son, Hyeon-Dong",
    title = "{Second Gegenbauer moment of a $\rho$-meson distribution amplitude}",
    eprint = "2008.06270",
    archivePrefix = "arXiv",
    primaryClass = "hep-ph",
    doi = "10.1103/PhysRevD.102.114005",
    journal = "Phys. Rev. D",
    volume = "102",
    number = "11",
    pages = "114005",
    year = "2020"
}
\end{document}